\documentclass[aps,prl,showpacs,twocolumn,superscriptaddress]{revtex4-1}
\usepackage{graphicx,amsmath,amssymb}
\usepackage[usenames]{color}
\usepackage{indentfirst}
\usepackage{float}
\usepackage{mathrsfs}
\begin{document}
\title{Universal scaling and critical exponents of the anisotropic quantum Rabi model}

\author{Maoxin Liu}
\affiliation{Beijing Computational Science Research Center, Beijing 100084, China}

\author{Stefano Chesi}
\email{stefano.chesi@csrc.ac.cn}
\affiliation{Beijing Computational Science Research Center, Beijing 100084, China}

\author{Zu-Jian Ying}
\affiliation{Beijing Computational Science Research Center, Beijing 100084, China}

\author{Xiaosong Chen}
\affiliation{Institute of Theoretical Physics, Chinese Academy of Sciences, P.O. Box 2735, Beijing 100190, China}
\affiliation{School of Physical Sciences, University of Chinese Academy of Sciences, Beijing 100049, China}

\author{Hong-Gang Luo}
\email{luohg@lzu.edu.cn}\affiliation{Center for Interdisciplinary Studies $\&$ Key Laboratory for
Magnetism and Magnetic Materials of the MoE, Lanzhou University, Lanzhou 730000, China}
\affiliation{Beijing Computational Science Research Center, Beijing 100084, China}

\author{Hai-Qing Lin}
\affiliation{Beijing Computational Science Research Center, Beijing 100084, China}


\begin{abstract}
We investigate first- and second-order quantum phase transitions of the anisotropic quantum Rabi model, in which the rotating- and counter-rotating terms are allowed to have different coupling strength. The model interpolates between two known limits with distinct universal properties. Through a combination of analytic and numerical approaches we extract the phase diagram, scaling functions, and critical exponents, which allows us to establish that the universality class at  finite anisotropy is the same as the isotropic limit. We also reveal other interesting features, including a superradiance-induced freezing of the effective mass and discontinuous scaling functions in the Jaynes-Cummings limit. Our findings are relevant in a variety of systems able to realize strong coupling between light and matter, such as circuit QED setups where a finite anisotropy appears quite naturally.
\end{abstract}

\maketitle

\textit{Introduction.--} While critical phenomena are traditionally associated with collective behavior in the thermodynamic limit, quantum phase transitions in systems with few degrees of freedom were recently brought to prominence \cite{Hwang2015, Hwang2016}. As it turns out, the topic is of great relevance for ongoing efforts on enhancing and  engineering light-matter interactions. By achieving the strong \cite{Wallraff2004,Devoret2007}, ultrastrong \cite{Bourassa2009, Niemczyk2010, Peropadre2010, Ridolfo2012, Todorov2014, Baust2016, Forn-Diaz2016}, and even deep strong coupling regime \cite{Casanova2010, Liberato2014, Fornd2017, Yoshihara2017, chenzhen2017}, atomic and solid-state resonances are able to induce profound modifications of the photon fields they interact with. The quantum Rabi model (QRM), describing a two-level system coupled to a single electromagnetic mode, represents the simplest realization of such light-matter interactions \cite{Rabi1936, Rabi1937}. Thus, it has served as a paradigmatic example to explore this kind of strong-coupling phenomena, and has received renewed attention in recent years \cite{Forn-Diaz2010, Braak2011, Solano2011, Bakemeier2012, Chen2012, Ashhab2013, Wolf2013, Ying2015, Hwang2015,  Braak2016}.

Surprisingly, an analytic solution of the QRM was only found recently, and has also motivated proposing a novel operational criterion of integrability \cite{Braak2011}. More directly related to the present study are several recent analyses on the dependence of QRM ground-state properties on the coupling strength  \cite{Ashhab2013, Ying2015, Hwang2015}.  An ansatz for the ground state based on the polaron and antipolaron concept was introduced in Ref.~\cite{Ying2015}, which has presented a phase diagram where the quadpolaron dominates in the weak coupling regime and the bipolaron dominates at strong coupling. The crossover between these two types of ground state becomes sharper by reducing the bosonic frequency \cite{Ashhab2013, Ying2015}, namely, in the classical oscillator limit \cite{Bakemeier2012}. It was proved that this behavior indeed reflects the existence of a true quantum phase transition (QPT), and the static and dynamical properties of the critical point were studied in detail  \cite{Hwang2015}. Furthermore, it was later found  that also the Jaynes-Cummings (JC) model \cite{Jaynes1963} exhibits a second-order QPT in the same limit \cite{Hwang2016}.

These findings have motivated us to study the physics of the QPT in the anisotropic QRM \cite{Xie2014}, which includes the two known limits as special cases. The model reads
\begin{equation}\label{Hamiltonian_Rabi}
H = \omega a^{\dagger}a+\frac{\Omega}{2}\sigma_x+g[(\sigma_+ a + \sigma_-a^{\dagger})+\lambda(\sigma_+a^{\dagger}+\sigma_-a)],
\end{equation}
where $\hbar = 1$, $a^\dagger$($a$) is the creation (annihilation) operator of the bosonic field oscillator, and $\boldsymbol{\sigma}$ are the Pauli matrices with $\sigma_{\pm}=\frac{1}{2}(\sigma_z \mp i\sigma_y)$. As known, the anisotropic QRM is relevant in a variety of systems including quantum well with spin-orbit
coupling \cite{Schliemann2003, Wangzh2016} and circuit QED, where strong interactions were realized \cite{Wallraff2004,Devoret2007,Niemczyk2010, Forn-Diaz2010, Forn-Diaz2016,Fornd2017,Yoshihara2017,chenzhen2017}. In this case, the asymmetry between rotating- and counter-rotating terms is often the typical scenario. For example, in the proposal of Ref.~\cite{Baksic2014} the degree of anisotropy is simply given by the relative strength of the inductive and capacitive couplings to the cavity. Also in a setup of two coupled SQUIDs, which was not specifically designed to implement Eq.~(\ref{Hamiltonian_Rabi}), measurements of the Bloch-Siegert shift are in good agreement with an intermediate value $\lambda \simeq 0.5$~\cite{Forn-Diaz2010,Xie2014}. The anisotropic QRM could also be implemented with trapped-ions \cite{Puebla2016} and is equivalent to a Hamiltonian of spin-orbit coupled electrons in semiconductors \cite{Xie2014} (possibly emulated by fermionic gases).

From the theoretical perspective, Eq.~(\ref{Hamiltonian_Rabi}) allows us to address systematically the role of the counter-rotating terms in the QPT, since the parameter $\lambda$ can interpolate between the QRM ($\lambda=1$) and JC model ($\lambda=0$).  The intermediate case is of special interest because the two known limits have different critical exponents and types of broken symmetry phase \cite{Hwang2015, Hwang2016}. Thus, the question about intermediate values of $\lambda$ arises very naturally. In the following, we develop a unified treatment which allows us to identify the appropriate scaling parameters and establish the complete phase diagram in the presence of  anisotropy. We find that the QPT at fixed $\lambda$ exhibits a remarkable degree of universality: it occurs between a normal and superradiant  (x- or p-type) phase, with critical exponents and scaling functions which are independent of $\lambda \neq 0$. Therefore, the generic realization of the QPT belongs to a well-defined universality class, which is the same of $\lambda = 1$ limit. On the other hand, by extending previous work on the JC model \cite{Hwang2016}, we expose the singular character of the QPT at $\lambda=0$.

\begin{figure}
\includegraphics[width=0.98\columnwidth]{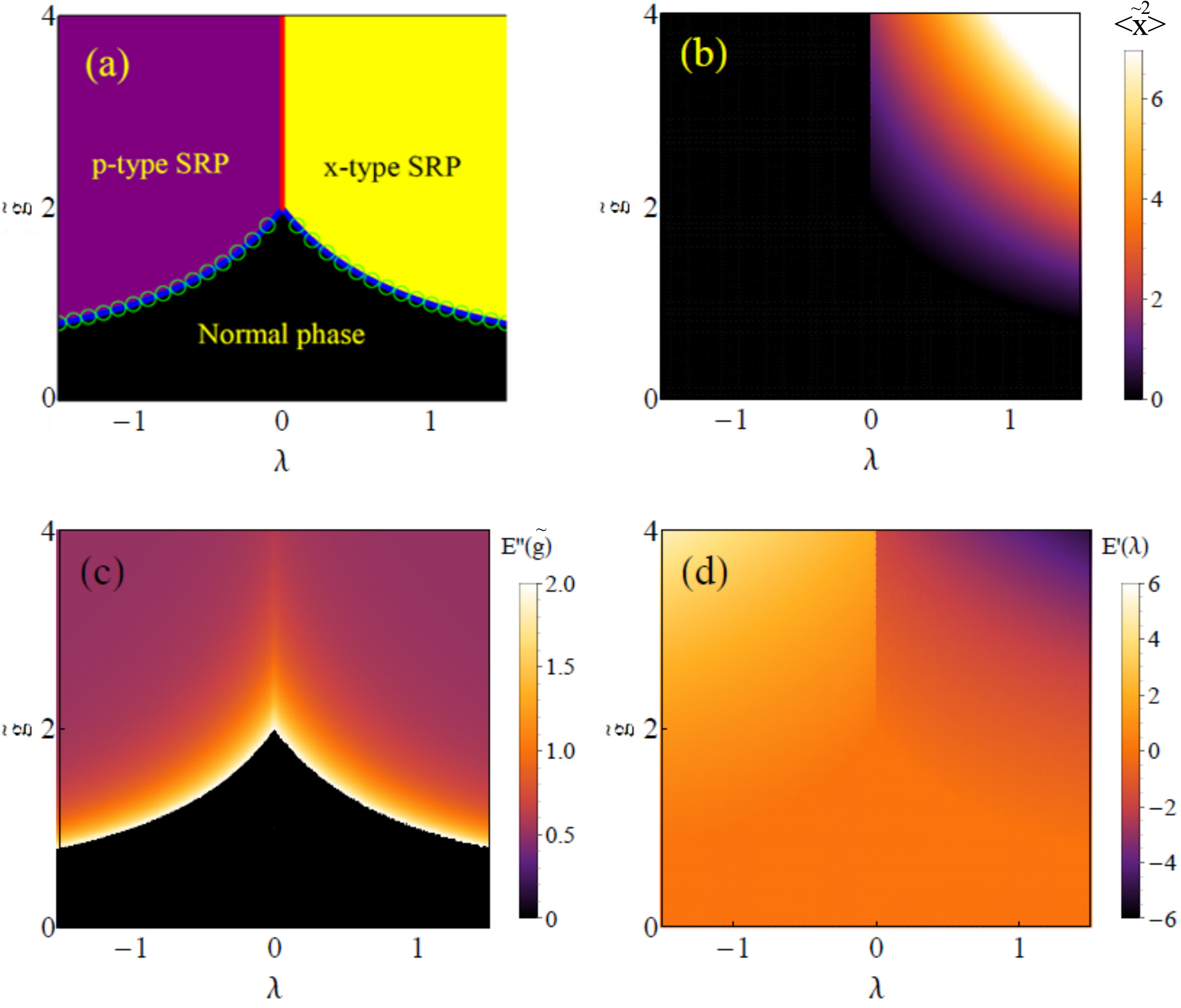}
\caption{(Color online) Properties of the QRM ground state in the $\eta \to \infty$ limit. (a): Phase diagram. The blue (red) line indicates a second-order (first-order) quantum phase transition. The green circles are obtained numerically. The other three panels show the following quantities: (b): $\langle \tilde{x}^2\rangle $ (where $\tilde x= x/\sqrt{\eta}$);  $\langle p^2 \rangle/\eta$ is available applying $\lambda \to -\lambda$.  (c): $\partial ^2 \tilde{E}_{gs}/\partial \xi^2$; (d): $\partial \tilde{E}_{gs}/\partial \lambda $.  }
\label{fig1}
\end{figure}

\textit{Phase diagram.--} To understand the main features of the phase diagram, we first rewrite Eq.~(\ref{Hamiltonian_Rabi}) by using dimensionless coordinate and momentum operators, $x=\frac{1}{\sqrt{2}}(a^{\dagger} + a)$ and $p=\frac{i}{\sqrt{2}}(a^{\dagger}-a)$. Within a constant term:
\begin{equation}\label{xp_dis}
H= \frac{p^2 + x^2}{2\eta}+\frac{\sigma_x}{2} + \tilde{g}\left[\frac{1+\lambda}{\sqrt{8\eta}}\sigma_z x+\frac{1-\lambda}{\sqrt{8\eta}}\sigma_y p\right],
\end{equation}
where $\eta = \Omega/\omega$ and $\tilde{g}=g/g_{c,0}$, with $g_{c,0} = \sqrt{\omega\Omega}/2$ \cite{Ashhab2013}. To simplify the notation, we set $\Omega=1$ in Eq.~(\ref{xp_dis}) and hereafter. First of all, we note that the second term becomes dominant in the $\eta \to \infty$ limit of interest, thus the relevant low-energy states have $\langle\sigma_x \rangle \simeq -1 $. Within this subspace, the ground-state is determined by the competition between the first term (a conventional oscillator) and the last term (the coupling between the bosonic mode and the two-level system). The coupling term has a larger prefactor, proportional to $\eta^{-1/2}$, but is off-diagonal in $\sigma_x$. Treating it with second-order perturbation theory gives:
\begin{equation}\label{xp_eff_2order}
H_{eff} \simeq \frac{p^2 + x^2}{2\eta} - \tilde{g}^2\frac{(1+\lambda)^2 x^2+(1-\lambda)^2 p^2}{8\eta}+ \ldots ,
\end{equation}
which shows that the oscillator term is dominant in the weak coupling regime, corresponding to the normal phase. On the other hand, at sufficiently large $\tilde g$ the coupling term will dominate and the Hamiltonian in Eq.~(\ref{xp_eff_2order}) becomes unbounded. To this order of approximation, $H_{eff}$ implies divergent values of $\langle x^2 \rangle,\langle p^2 \rangle$, which in turn signals the onset of a superradiant phase.  By introducing $\xi=\tilde{g}(1+\lambda)/2$ and $\xi'=\tilde{g}(1-\lambda)/2$, the gap $\varepsilon/\eta$ of the normal phase is simply given by:
\begin{equation}\label{epsilon_gap}
\varepsilon = \sqrt{(1-\xi^2)(1-{\xi'}^2)},
\end{equation}
which becomes zero at $\tilde{g}_c = \frac{2}{1+|\lambda |}$. This phase boundary is plotted in Fig.~\ref{fig1}(a) as a solid (blue) line and recovers the known results for the isotropic QRM \cite{Ashhab2013, Hwang2015} and JC model \cite{Hwang2016} at $\lambda = 1$ and $\lambda = 0$, respectively. The onset of instability in  Eq.~(\ref{xp_eff_2order}) is due to the $x^2$ ($p^2$) terms when $\lambda > 0$ ($\lambda < 0$), suggesting that the superradiant phase should be divided into two regimes according to the sign of the anisotropy parameter $\lambda$. This observation is reflected by the vertical (red) phase boundary of Fig.~\ref{fig1}(a). All these conclusions are further confirmed by the numerical analysis (see Fig. 1) and will be more rigorously justified and extended  in the rest of the paper.

\textit{Classical oscillator limit.--} For $\lambda > 0$, the $\sigma_z x$ coupling plays a dominant role and, to proceed, it is helpful to rescale the coordinate by $\tilde {x} = x / \sqrt{\eta}$. Correspondingly, $\tilde{p} = -i\partial/\partial \tilde{x}$. Thus, $H$ becomes
\begin{equation}\label{xp_dis2}
\tilde{H} = \frac{\tilde{p}^2}{2\eta^2} + \frac{1}{2}\tilde{x}^2 + \frac{1}{2}\sigma_x + \frac{\xi}{\sqrt{2}}\sigma_z \tilde{x}+\frac{\xi'}{\sqrt{2\eta}}\sigma_y \tilde{p}.
\end{equation}
By taking the classical oscillator limit \cite{Bakemeier2012}, i.e., $\eta \rightarrow \infty$ (but keeping $\tilde{g}$ finite), one can drop the first and last terms of Eq.~(\ref{xp_dis2}).
The remaining Hamiltonian is readily diagonalized and has eigenvalues
\begin{equation} \label{gs-energy}
\tilde{E}_{\pm} = \frac{1}{2}\left(\tilde{x}^2 \pm \sqrt{1 + 2\xi^2 \tilde{x}^2}\right),
\end{equation}
where $\tilde x$ is now a classical coordinate. The lower branch $\tilde{E}_{-}$ has the standard behaviour of the Landau potential across a continuous phase transition, where $\xi$ plays a role analogous to the inverse temperature: $\tilde{E}_{-}$ has one minimum around $\tilde{x}=0$ when $\xi<1$ and two minima at finite $\tilde{x}$ when $\xi > 1$.  The order parameter is:
\begin{equation}\label{expy}
\tilde{x}_{0,\pm} =
\pm \sqrt{\frac{\xi^2-\xi^{-2}}{2}} ~ \theta(\xi-1),
\end{equation}
where $\theta(x)$ is the step function and the critical value $\xi_c=1$ is in agreement with the $\lambda>0$ side of Fig.~\ref{fig1}.  The ground state energy $\tilde{E}_{gs} = -\frac{1}{2} -\frac{1}{4} \left(\xi^2 + \xi^{-2}-2\right)\theta(\xi-1)$ is easily obtained from Eqs~(\ref{xp_dis2}) and (\ref{expy}) and indicates a second-order QPT at $\xi_c$, since the first-order derivative $\partial\tilde{E}_{gs}/\partial \xi$ is continuous but the second-order derivative  $\partial^2\tilde{E}_{gs}\partial \xi^2$ is not [see Fig.~\ref{fig1}(c)]. We also emphasize that, although the anisotropy parameter $\lambda$ obviously influences the critical coupling strength $\tilde{g}_c$ as well as $\tilde{x}_{0,\pm}$ and $\tilde{E}_{gs}$, the functional dependence of these physical quantities becomes universal -- in the sense of being independent of $\lambda$ -- once the problem is formulated in terms of the rescaled coupling $\xi$.

The other physical quantities have also been calculated~\cite{Suppl_Info}. In particular, $\langle\tilde{x}^2\rangle = \tilde{x}_{0,\pm}^2=(\xi^2-\xi^{-2})/2$ and the correlation function $\langle \sigma_z \tilde{x}\rangle = -(\xi - \xi^{-3})/\sqrt{2}$ have finite values as the system enters into the superradiant phase. On the other hand, $\langle\tilde{p}^2\rangle/\eta^2 = \frac{1+\lambda}{4\eta\sqrt{\lambda}}\sqrt{1-\xi^{-4}}$ and $\langle \tilde{p} \sigma_y \rangle /\eta = \frac{\xi^{-3}}{\sqrt{2}\eta}  \left( 1- \frac{1-\lambda}{2\sqrt{\lambda}}\sqrt{\xi^4-1}\right)$ tend to zero as $\eta\rightarrow \infty$, which is compatible with dropping terms involving the momentum in the classical oscillator limit.

The case $\lambda < 0$ can be treated in a similar way, or by using the fact that Eq.~(\ref{xp_dis}) remains unchanged under the following  mapping
$\mathscr{F}=\mathscr{F}_1\otimes \mathscr{F}_2$, which contains a $Z_2$ symmetric mapping $\mathscr{F}_1$ : $\{H(\lambda) \rightarrow H(-\lambda)\}$, and a unitary transformation $\mathscr{F}_2$ : $\{ H \rightarrow U^{\dag}HU \}$, where $U=e^{-i\frac{\pi}{2}a^{\dag}a}\otimes e^{-i\frac{\pi}{4}\sigma_x}$.  The Hamiltonian in Eq.~\eqref{xp_dis} is a fixed point of the functional $\mathscr{F}$. Therefore, the properties we have discussed so far are readily translated to the $\lambda<0$ side of the phase diagram. In particular, the critical value $\xi'=1$ gives the phase boundary of Fig.~\ref{fig1} for $\lambda < 0$ (blue solid line). Furthermore, as made clear by the exact mapping, the superradiant phase at $\lambda<0$ is caused by the momentum $p$ rather than the position $x$. Thus, one gets two types of superradiant phases (x-type and p-type) and a sharp jump of both $\langle x^2 \rangle$ and $\langle p^2 \rangle$ at $\lambda=0$. One can also check that the first derivative of $\tilde{E}_{gs}$ is discontinuous at $\lambda=0$, since $\partial \tilde{E}_{gs}/\partial \lambda  |_{\lambda\rightarrow 0^\pm}  = \mp \left(\xi - 1/\xi^3\right) \tilde{g}/4$  [see Fig.~\ref{fig1}(d)]. This indicates the existence  in the superradiant regime of a first-order QPT dependent on the sign of $\lambda$. We stress that this new transition is only revealed by considering the complete phase diagram of the QRM, extended to include anisotropy.

\textit{Effective Hamilltonians.--} We would like next to address the critical scaling at the second-order QPT.
It begins with our analytic approach which leads to the following effective Hamiltonian \cite{Suppl_Info}:
\begin{equation}\label{effH2}
H_{eff} \simeq \frac{p^{2}+x^2}{2\eta}- \sqrt{\frac{1}{4} +\frac{\xi^2 x^2 +{\xi'}^2 p^2 -\xi\xi'}{2\eta} }.
\end{equation}
The above expression was obtained by performing an exact resummation of leading perturbative terms \cite{Suppl_Info}.
As we will see shortly, $H_{eff}$ is fully consistent with our previous discussion and, in fact, represents a generalization of  Eq.~(\ref{xp_eff_2order}) and the classical oscillator limit.

Firstly, we note that the order parameter in the superradiant phase ($\langle x^2 \rangle$ or $\langle p^2 \rangle$) is $\propto \eta$, thus an expansion of the square root in Eq.~(\ref{effH2}) becomes justified. Focusing on the x-type phase, we can neglect to first approximation $p^2$ and $\xi \xi'$, to recover the classical potential $\tilde{E}_-$ of Eq.~(\ref{gs-energy}). If higher orders in $\eta^{-1}$ are considered  we obtain the following approximation (in rescaled coordinates), which is suitable to characterize the scaling properties at $\xi \simeq 1$:
\begin{equation}\label{effH}
\tilde{H}_{eff}\simeq -\frac{1}{2}+\frac{\xi\xi'}{2\eta} +\frac{2 \lambda}{(1+\lambda)^2} \frac{\tilde{p}^2}{\eta^2}+\frac{1-\xi^2}{2 } \tilde{x}^2+\frac{\xi^4}{4\eta}\tilde{x}^4.
\end{equation}
Here, the quartic potential is simply the small-x expansion of $\tilde{E}_-$ [see Eq.~(\ref{gs-energy})] and is valid for $\langle \tilde x^2\rangle \ll \eta |1-\xi | $ (i.e., $\xi\to 1$). The kinetic term was also derived under the asumption $\xi=1$.  Since $\langle x^2 \rangle$, $\langle p^2 \rangle$ are small in the critical regime, low-order perturbation theory becomes accurate and Eq.~(\ref{effH}) can be confirmed by a direct calculation. To do that, we have chosen $H_0 =\sigma_x/2+ (p^2+x^2)/(2\eta)$ as unperturbed Hamiltonian and $V_{od} = (\xi x\sigma_z + \xi' p\sigma_y)/\sqrt{2\eta}$ as off-diagonal perturbation and computed the Schrieffer-Wolff transformation up to fourth-order \cite{Suppl_Info}.

Equation~(\ref{effH}) also defines the effective mass:
\begin{equation}\label{Meff}
M_\lambda = \eta^2 \frac{(1+\lambda)^2}{4\lambda},
\end{equation}
which plays a central role for the critical scaling and the stability of the superradiant phase. To extend Eq.~(\ref{Meff}) beyond $\xi\simeq 1$, one can derive from Eq.~(\ref{effH2}) a kinetic term of the form $\frac{\tilde{p}^2}{2\eta^2}\left(1-{\xi'}^2/\sqrt{1+2\xi^2 \tilde x_{0,\pm}^2}\right)$, which is in agreement with Eq.~(\ref{xp_eff_2order}) if $\xi <1 $ (i.e., $\tilde{x}_{0,\pm}=0$). This expression implies that fluctuations of $p$ are promoted by the interaction in the normal regime, where a larger $\tilde g$ enhances the effective mass. In the superradiant phase, however, the second-order terms of Eq.~(\ref{xp_eff_2order}) give an incorrect result and, in particular, naively predict a phase transition at $\xi'=1$. This instability is prevented by the finite value of $\tilde x_{0,\pm}$, which takes into account non-perturbative effects beyond second-order. Using Eq.~(\ref{expy}) leads to the remarkable result of a \emph{constant} effective mass  in the superradiant phase: Although the mass enhancement is actually due to the interaction, it becomes independent on the coupling strength for $\xi>1$ and only reflects the interaction anisotropy. In other words, the formation of the x-type superradiant phase freezes the effective mass renormalization at the $\xi=1$ value.

As a final remark on Eq.~(\ref{Meff}) we note that in the isotropic QRM there is no renormalization effect  ($M_{\lambda=1} = \eta^2$). Thus, the peculiar interplay of superradiance and effective mass renormalization between the two quadratures is specifically related to the intermediate values of $\lambda$ considered here.

\begin{figure}
\includegraphics[width=0.48\columnwidth]{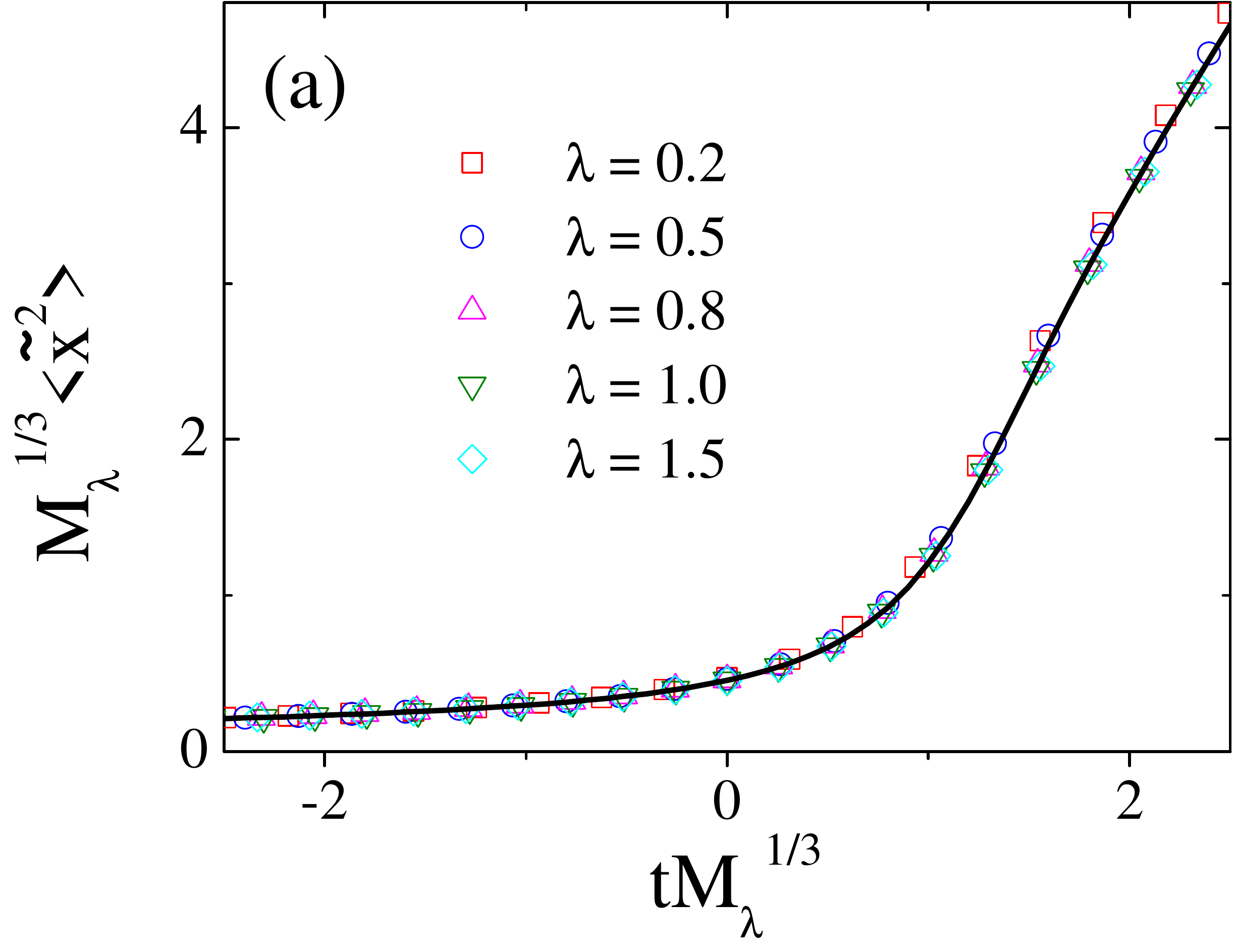}
\includegraphics[width=0.48\columnwidth]{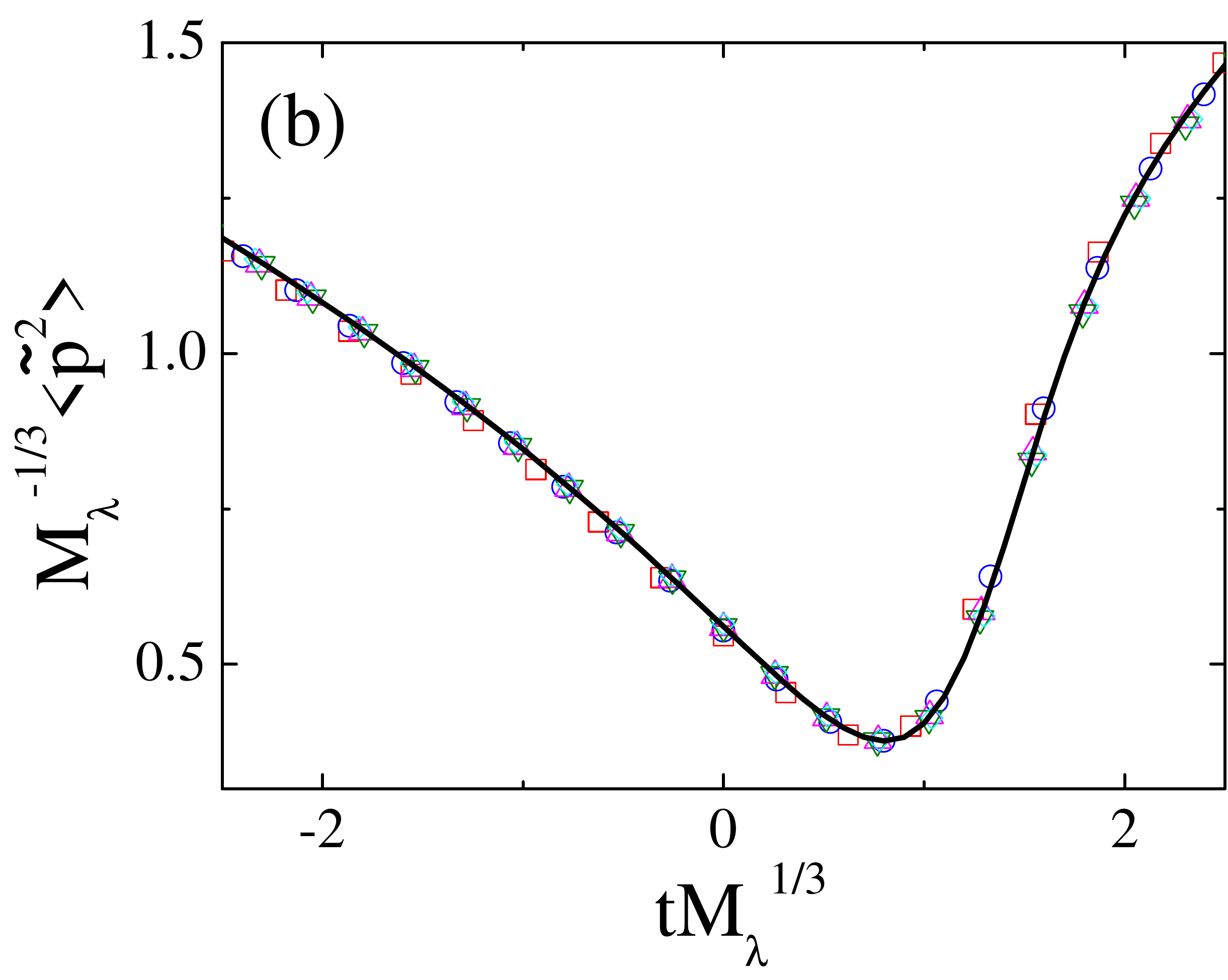}
\caption{(Color online) (a): Universal scaling function for $\langle \tilde{x}^2\rangle$.  Symbols are obtained by a solution of the full anisotropic QRM, with $\eta = 2^{20}$ and different values of $\lambda$. All numerical data collapse into the single function $X_1(v)$ (solid curve), obtained from Eq.~(\ref{phi_SE}). (b): The same analysis applied on $\langle \tilde{p}^2\rangle$ confirms the universal scaling law $P_1(v)$ (solid curve).}
\label{fig2}
\end{figure}

\textit{Finite-$\eta$ scaling.--} We are now ready to discuss the scaling properties, based on Eq.~(\ref{effH}) and the definition of $M_\lambda$. To reveal the universal properties of the phase transition, it is appropriate to introduce the scaling variables $u=\tilde{x}M_{\lambda}^{1/6}$ and $v=t M_\lambda^{1/3}$ (where $t=\xi-1$). Although the ground state wavefunction $\phi_0$ depends in general on four variables ($t,~x,~\eta,$ and $\lambda$), it is described by a simple equation in terms of $u,v$:
\begin{equation}\label{phi_SE}
\left(-\frac{1}{2}\frac{\partial ^2}{\partial u^2}-vu^2+\frac{u^4}{4}\right)\phi_0(u,v) =E_0(v) \phi_0(u,v),
\end{equation}
where $E_0(v)$ gives the ground-state energy:
\begin{equation}\label{E_correction}
E_G(\lambda)=-\frac{1}{2}+\frac{1}{2\eta}\frac{1-\lambda}{1+\lambda}+M_{\lambda}^{-2/3}E_{0}(t M_\lambda^{1/3})+\ldots.
\end{equation}
The scaling law obeyed by a certain observable is easily derived from the scaling form  $\phi_0(\tilde{x} M _{\lambda}^{1/6} ,tM_{\lambda}^{1/3})$ of the ground state. In the important case of $x^{2n},p^{2n}$ we obtain:
\begin{align}\label{scaling_x0}
&\langle x^{2n}\rangle = \left(\eta M_{\lambda}^{-1/3}\right)^{n}X_n\left(tM_{\lambda}^{1/3}\right),\\
\label{scaling_p0}
&\langle p^{2n}\rangle = \left(\eta M_{\lambda}^{-1/3}\right)^{-n}P_n\left(tM_{\lambda}^{1/3}\right),
\end{align}
where the universal functions $X_n(v),P_n(v)$ are given by expectation values over $\phi_0(u,v)$ and can be readily evaluated \cite{Suppl_Info}. We have confirmed the validity of our treatment by direct numerical solution of $H$ [see Eq.~(\ref{Hamiltonian_Rabi})]. Figure~\ref{fig2} shows that the numerical values of $ \langle x^2 \rangle$ and $ \langle p^2 \rangle$ at large $\eta$ and different values of $\lambda$ all collapse into a single curve when appropriately scaled. The two numerical scaling functions agree with the $X_1(v),P_1(v)$ obtained from Eq.~(\ref{phi_SE}). We thus conclude that the presence of anisotropy does not modify either the critical exponents or the scaling behavior and identify the whole second-order phase transition line as belonging to the same universality class \cite{Jonas2016}. On the other hand, the JC model ($\lambda=0$) is a special case which will be discussed next.

\begin{figure}
\begin{centering}
\raisebox{2pt}{\includegraphics[width=0.51\columnwidth]{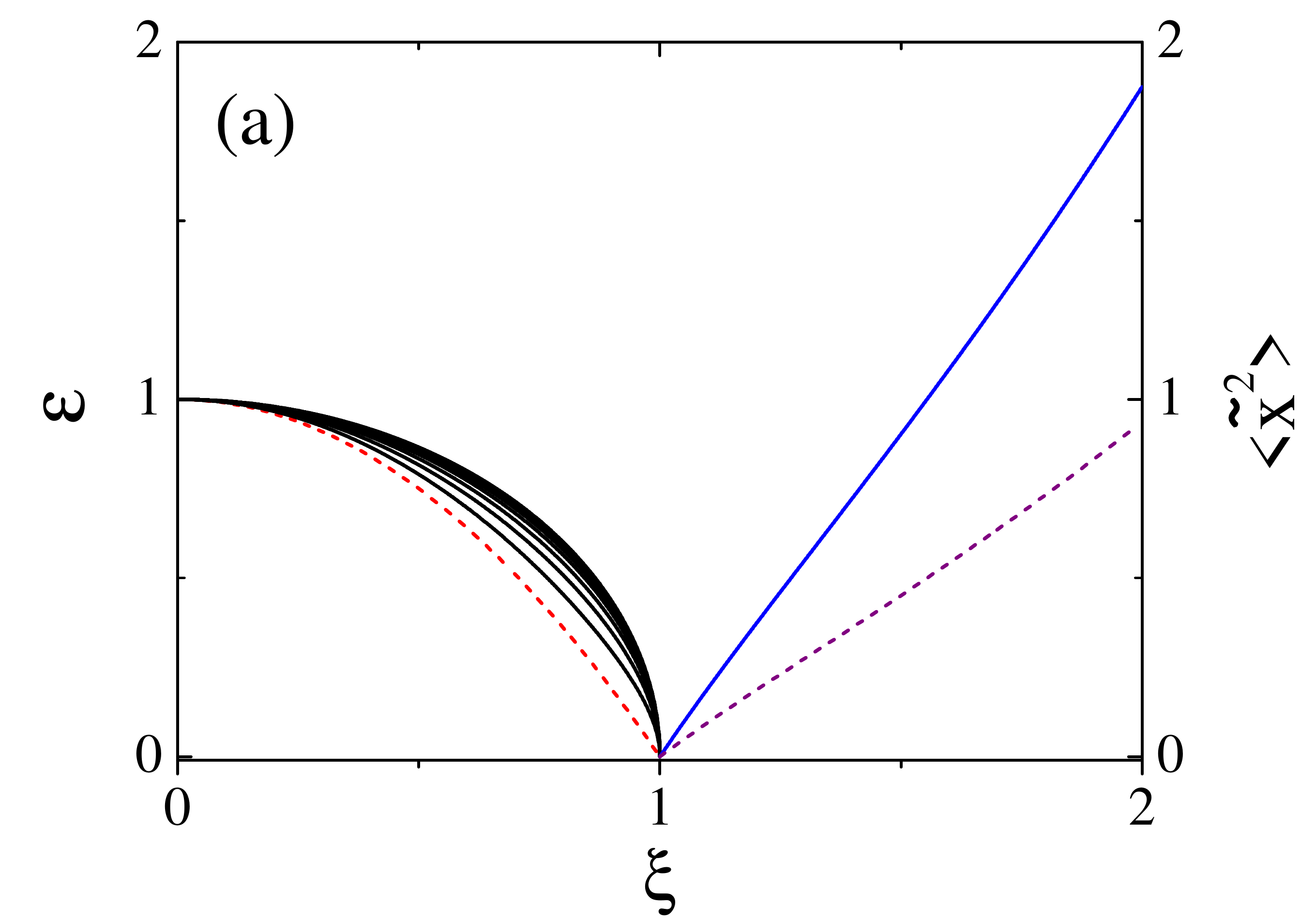}}
\raisebox{0pt}{\includegraphics[width=0.47\columnwidth]{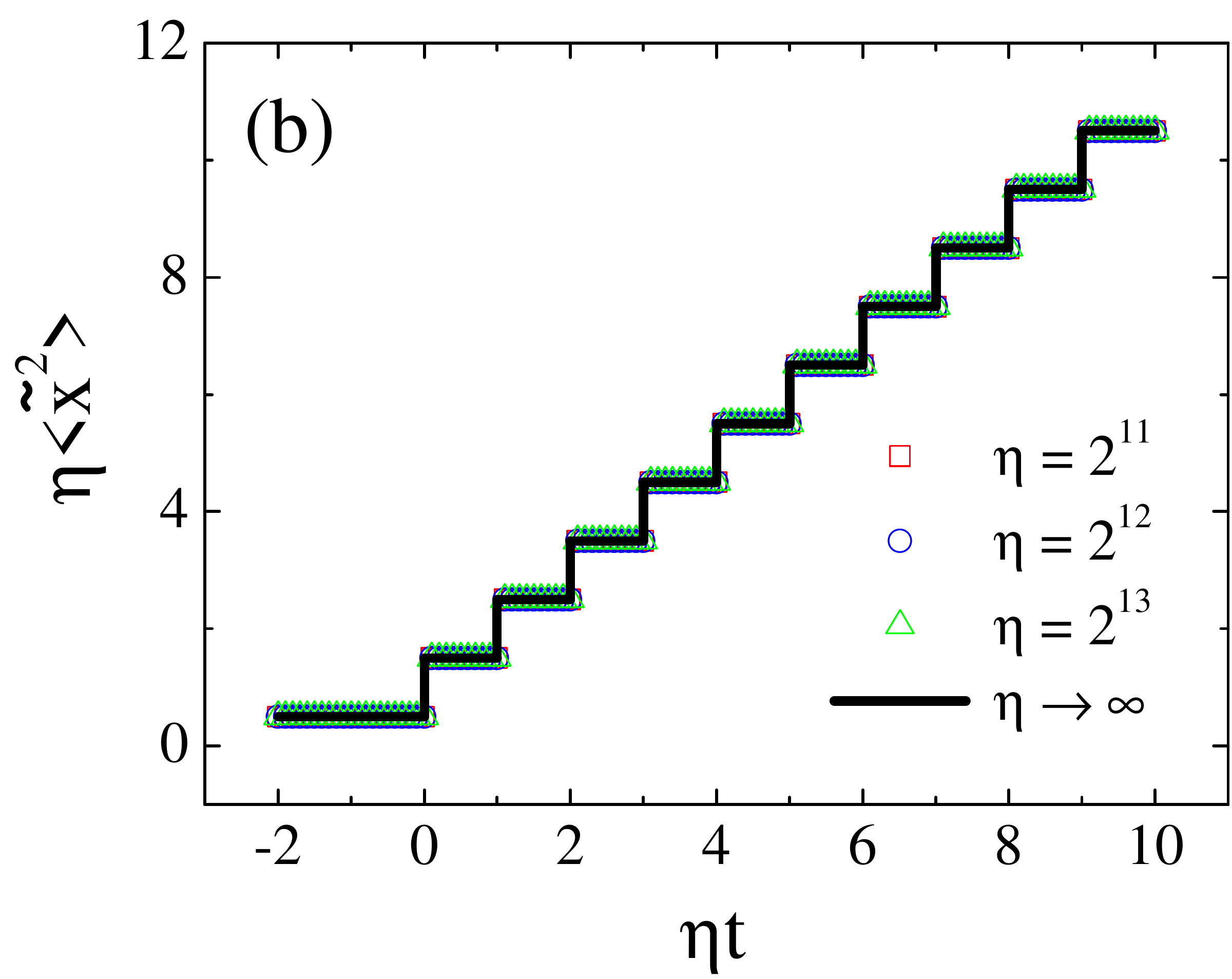}}
\vspace{-0.3cm}
\end{centering}
\caption{(Color online)
(a)  For $\xi<1$ we plot the energy gap in the normal phase, given by Eq.~(\ref{epsilon_gap}). Solid curves are for $\lambda=0.1,0.2,\ldots 1$ (bottom to top) and the dashed curve is for $\lambda=0$. Note the difference in critical exponents. For $\xi>1$ we plot $\langle \tilde{x}^2 \rangle$. The solid curve is valid for all the $\lambda >0$ while the dashed curve is the anomalous $\lambda=0$ result.  (b): scaling function for $\langle x^2 \rangle=\langle p^2 \rangle$ and $\lambda=0$. The numerical values at large $\eta$ (symbols) are in agreement with Eq.~(\ref{scalingJC}) (solid line).
}
\label{fig3}
\end{figure}

\textit{The JC critical line.--}  Due to the presence of the first-order transition at $\lambda=0$, the scaling functions have an abrupt change  on that critical line. Since the exact mapping between opposite values of $\lambda$ interchanges the roles of $x$ and $p$, the right-hand sides of Eqs.~(\ref{scaling_x0}) and (\ref{scaling_p0}) must be switched when $\lambda<0$ (together with the changes $M_\lambda \to M_{-\lambda}$ and $t \to \xi'-1$).  At $\lambda=0$ all the quadratures are equivalent since $\xi=\xi'$ and Eq.~(\ref{effH2}) becomes a function of $p^2+x^2$. Therefore, we recover the Mexican-hat potential in phase space discussed in Ref.~\cite{Hwang2016}, together with the related spontaneous breaking of $U(1)$ symmetry and gapless Goldstone mode.

The anomalous nature of the JC model in the phase diagram is reflected by other critical properties. For example,  at $\lambda=0$ the critical exponent $\alpha$ for the excitation gap is different from the rest of the phase diagram ($\alpha=1$ instead of $1/2$). The different value of $\alpha$ in the QRM and JC model was noted already in Ref.~\cite{Hwang2016}, but our analysis is much more general as it extends the $\alpha = 1/2$ exponent to all values $\lambda\neq 0$. Furthermore, while our Eq.~(\ref{expy}) is universally applicable for $\lambda > 0$, we find that the corresponding result at $\lambda=0$ is half of Eq.~(\ref{expy}). These two properties are illustrated by Fig.~\ref{fig3}(a), where the energy gap ($\xi<1$) and the order parameter ($\xi>1$) are plotted.

Finally, we have also considered the scaling functions of the JC model. The expectation values of  $\langle x^{2n} \rangle = \langle p^{2n} \rangle$ in the $\eta \to \infty$ limit are given by:
\begin{equation}\label{scalingJC}
\frac{(2n-1)!!}{2^n}\left[ 1+\sum_{q=0}^{\infty} \theta(\eta t-q)(D(n,q+1)-D(n,q)) \right],
\end{equation}
where $D(n,q)$ are the Delannoy numbers \cite{Suppl_Info}. The case $n=1$ is shown in Fig.~\ref{fig3}(b) and should be compared to Fig.~\ref{fig2}. We see that $\eta$ plays the role of $M_\lambda^{1/3}$ (which is not defined at $\lambda = 0$) and that the functional dependence is completely different from the rest of the phase diagram. In particular, the scaling function has a discontinuous nature. This reflects the fact that the QPT of the JC model is given by a succession of level crossings \cite{Hwang2016,Suppl_Info} (as seen, the spacing is $\Delta \xi \simeq 1/\eta$). Such scaling functions highlight again the singular nature of the $\lambda=0$ line within the phase diagram.

\textit{Conclusion.--} We have characterized the QPTs of the QRM as a function of coupling strength and anisotropy and established the universal character of the second-order phase transition for $\lambda \neq 0$. A first-order critical line at $\lambda=0$ separates two types of superradiant phases. Besides universality, we have found other interesting features such as the freezing of the effective mass (induced by the broken-symmetry in $x$) and the discontinuous scaling functions of the JC model. Our results emphasize the critical role played by counter-rotating terms, whose current experimental relevance is due to the rapid progress in enhancing light-matter interactions \cite{Wallraff2004,Devoret2007,Bourassa2009, Niemczyk2010, Peropadre2010, Ridolfo2012, Todorov2014, Baust2016, Forn-Diaz2016,Casanova2010, Liberato2014,Fornd2017, Yoshihara2017,chenzhen2017,Groblacher2009,Gunter2009,Niemczyk2010}. In particular, the exposed singularity of the JC limit implies that even tiny counter-rotating terms lead to dramatic changes of the scaling behavior. Since a finite anisotropy is quite natural \cite{Forn-Diaz2010,Xie2014}, the universal scaling features discussed here may be tested by practical implementations.
A study of the universality from the point of view of critical dynamics would be another interesting extension of this work.
Finally, we note that the superradiant phase leads to a strongly squeezed ground state of light, which has potential value for metrology and enhanced sensing applications.

\textit{Acknowledgment.--} This work is supported by NSFC (Grants No. 11604009, No. 11574025, No. 11325417, No. 11674139, No. 1121403), NSAF (Grant No. U1530401). SC acknowledges support from the National Key Research and Development Program of China (Grant No. 2016YFA0301200). We also acknowledge the computational support in Beijing Computational Science Research Center (CSRC).

\appendix
\begin{widetext}
\begin{center}
\textbf{\large Supplemental Material for ``Universal scaling and critical exponents of the anisotropic
quantum Rabi model"}
\end{center}
\section{Effective Hamiltonians}

We discuss here a perturbative treatment of the anisotropic QRM [see Eq.(2) in the main text]. We first consider the Sch rieffer-Wolff (SW) transformation, which provides a general method to derive low-energy effective Hamiltonians. The lower orders are relatively straightforward to  evaluate and results up to fourth-order are given below, together with some details of the derivation. To extend such low-order SW Hamiltonian, an exact resummation of the leading contributions can be performed, which is described later in this Section.

\subsection{SW effective Hamiltonian}

For convenience, we give immediately the final result of the SW transformation (with $\Omega=1$):
\begin{eqnarray}\label{Heff_SW}
H_{eff} &\simeq &-\frac{1}{2}+\frac{p^2+x^{2}}{2\eta} \nonumber \\
&&-\frac{\tilde{g}^{2}}{8}\frac{\eta}{\eta ^{2}-1}\left[
(1+\lambda )^{2}x^{2}+(1-\lambda )^{2}p^{2}-1+\lambda ^{2}+\frac{1-\lambda
^{2}}{\eta }\left( x^{2}+p^{2}\right) -\frac{1+\lambda ^{2}}{\eta}\right] \nonumber \\
&&+\frac{\tilde{g}^{4}}{64\eta^2 }\left[ \left( 1+\lambda \right)
^{2}x^{2}+\left( 1-\lambda \right) ^{2}p^{2}-\left( 1-\lambda ^{2}\right) %
\right] ^{2}+\ldots
\end{eqnarray}
where the first line is the unperturbed Hamiltonian in the $\sigma_x =-1$ subspace. The second line is the full second-order result and the third line is the leading contribution from the fourth-order term (odd terms are all identically zero). It is easily checked that the first two lines recover Eq.~(3) of the main text, i.e., omitting unnecessary constants and considering the limit of large $\eta$.

In discussing the derivation of Eq.~(\ref{Heff_SW}), we refer to the brief summary of the SW transformation given in Appendix A of Ref.~\cite{SC}. In particular, we adopt the same notation. To apply the SW transformation, we define a low-energy subspace by introducing the following projectors:
\begin{equation}
P=\left\vert -\right\rangle \left\langle -\right\vert, \qquad
Q=\left\vert +\right\rangle \left\langle +\right\vert,
\end{equation}
where $|\pm\rangle$ are eigenstates of $\sigma_x$. We partition the anisotropic QRM into an unperturbed Hamiltonian and off-diagonal perturbation as follows:
\begin{equation}
H_{0}=\frac{1}{2}\sigma _{x}+\frac{p^{2}+x^2}{2
\eta},  \qquad
V_{od} =\tilde{g}\left(\frac{1+\lambda }{\sqrt{8\eta}}x\sigma _{z}+\frac{1-\lambda }{\sqrt{8\eta}}p\sigma _{y}\right).
\end{equation}
In general, the SW transformation is formulated in the presence of a diagonal perturabtion $V_d$ which, however,  is zero in our case. This simplifies things and, in particular, the first-order correction $H_{eff}^{(1)}=PV_{d}P=0$ is absent.

The second-order term is given as follows, in terms of the $L_0$ superoperator (defined by $L_0 A =[H_0, A]$):
\begin{equation}\label{H2_SW}
H_{eff}^{(2)}=\frac{1}{2}P[S_{1},V_{od}]P=\frac{1}{2}%
P[L_{0}^{-1}V_{od},V_{od}]P,
\end{equation}
and can be evaluated by a straightforward calculation. The following quantity is useful:
\begin{align}\label{S1_def}
S_{1} & =L_{0}^{-1}V_{od}=-i\lim_{s\rightarrow 0}\int_{0}^{\infty
}e^{-st}e^{iH_{0}t}V_{od}e^{-iH_{0}t}dt \nonumber \\
& =-i\tilde{g}\lim_{s\rightarrow
0}\int_{0}^{\infty }e^{-st}\left[ \frac{1+\lambda}{\sqrt{8\eta}}x(t)\sigma _{z}(t)+\frac{1-\lambda}{\sqrt{8\eta}}p(t)\sigma _{y}(t)\right] dt
\end{align}%
where
\begin{equation} \label{xpsigma}
\begin{array}{l}
x(t) =x\cos (t/\eta)+p\sin (t/\eta), \vspace{0.1cm} \\
p(t) =p\cos (t/\eta)-x\sin (t/\eta),
\end{array} \qquad\quad
\begin{array}{l}
\sigma _{z}(t) =\sigma _{z}\cos t+\sigma _{y}\sin  t, \vspace{0.1cm}\\
 \sigma _{y}(t) =\sigma _{y}\cos t-\sigma _{z}\sin t.
 \end{array}
 \end{equation}
After performing elementary time integrations,  Eqs.~(\ref{H2_SW}) and (\ref{S1_def}) yield the second line of Eq.~(\ref{Heff_SW}).

Since the third-order term $H_{eff}^{(3)}=\frac{1}{2}P[L_{0}^{-1}[S_{1},V_{d}],V_{od}]P$ is trivially zero, we focus on the fourth-order terms:
 \begin{equation}\label{H4_def}
H_{eff}^{(4)}=-\frac{1}{6}P\left[ L_{0}^{-1}\left[ S_{1},\left[ S_{1},\left[ S_{1},H_{0}%
\right] \right] \right] ,V_{od}\right] P-\frac{1}{24}P\left[ S_{1},\left[
S_{1},\left[ S_{1},V_{od}\right] \right] \right] P.
\end{equation}
which can be computed in a similar way to the second-order term. However, the full result is more cumbersome and is not given here. Instead, we concentrate on an approximation which useful in the limit of large $\eta$ and can be extended to higher orders (see the next Section). As seen in Eq.~(\ref{xpsigma}), the time dependence of $x(t),p(t)$ is much slower than $\sigma_{y,z}(t)$. Thus,  in evaluating $S_1$ to leading order,  we can substitute the $t=0$ value  $x(t),p(t) \to x,p$ and obtain:
\begin{eqnarray}\label{S1}
S_{1} \simeq -i\tilde{g}\left( \frac{1+\lambda }{\sqrt{8 \eta}}x\sigma
_{y}-\frac{1-\lambda }{\sqrt{8 \eta}}p\sigma _{z}\right),
\end{eqnarray}%
which allows us to directly compute the second term of Eq.~(\ref{H4_def}). In the first term, there is an additional time integral induced by $L_0^{-1}$. Performing the same approximation on $x(t),p(t)$ discussed above, we obtain the last line of Eq.~(\ref{Heff_SW}) as final result.

\subsection{Infinite resummation of the leading-order terms}

By considering Eq.~(\ref{Heff_SW}), we notice that we can approximate the second-order term at large $\eta$ and obtain a simple  form for $H_{eff}$:
\begin{equation}\label{Heff_Vseries}
H_{eff} \simeq -\frac{1 }{2}+\frac{p^{2}+x^2}{2\eta}- V^2 + V^4 + \ldots
\end{equation}
where:
\begin{equation}
V^2 = \langle - |V_{\rm od}^2 |-\rangle  =\frac{ \tilde{g}^2}{8\eta} \left[(1+\lambda)^2 x^2 +(1-\lambda)^2 p^2 -(1-\lambda^2)\right].
\end{equation}

The origin of the series in power of $V^2$, appearing in Eq.~(\ref{Heff_Vseries}), is not difficult to understand if we consider the matrix elements of the perturbative corrections. For the second-order term:
\[
\langle m | H_{eff}^{(2)} | m^{\prime }\rangle =\frac{1}{2}\sum_{l} \langle m | V_{od} | l \rangle \langle l |V_{od} | m^{\prime} \rangle \left[ \frac{%
1}{E_{ml}}+\frac{1}{E_{m^{\prime }l}}\right],
\]%
where $m,m',l$ label eigenstates of $H_0$ with energies $E_{l},E_{m'},E_{l}$. For the energy denominators, we used the notation $E_{ml}=E_m - E_l$. Since $m,m' \in P$ and $l \in Q$, the two energy denominators are both equal to $-1+ O(1/\eta)$ and, neglecting small corrections, we have:
\begin{equation}
\langle m | H_{eff}^{(2)} | m^{\prime }\rangle \simeq -\sum_{l} \langle m | V_{od} | l \rangle \langle l |V_{od} | m^{\prime} \rangle = -\langle m | V_{od}^{2} | m^{\prime },\rangle
\end{equation}
implying that $H_{eff}^{(2)}\simeq -V_{od}^2$ in the low-energy subspace. The effective Hamiltonian entering Eq.~(\ref{Heff_Vseries}) is $\langle - | H_{eff}^{(2)} | - \rangle  \ $, which leads to the $-V^2$ correction in Eq.~(\ref{Heff_Vseries}).

Similarly, we can consider the formula for the the fourth-order matrix elements \cite{Winkler_book}:
\begin{align}
\langle m | H_{eff}^{(4)} | m^{\prime }\rangle =-\frac{1}{24}\sum_{l,l^{\prime },m^{\prime \prime}} &
 \langle m | V_{od} | l \rangle \langle l |V_{od} | m^{\prime\prime} \rangle \langle m^{\prime\prime} | V_{od} | l' \rangle \langle l' |V_{od} | m^{\prime} \rangle \left( \frac{8}{E_{ml}E_{ml'}E_{m''l'}}
  +\frac{8}{E_{m'l}E_{m'l'}E_{m''l'}}  \right .  \nonumber \\
&+\frac{4}{E_{ml'}E_{m''l}E_{ml}}+\frac{4}{E_{ml'}E_{m''l}E_{m''l'}} +\frac{4}{E_{m'l}E_{m''l'}E_{m^{\prime }l'}}+\frac{4}{E_{m'l}E_{m''l'}E_{m^{\prime\prime }l}}\nonumber \\
& -\frac{1}{E_{m^{\prime \prime }l}E_{m^{\prime \prime}l'}E_{ml}}-\frac{1}{E_{m^{\prime \prime }l}E_{m^{\prime \prime}l'}E_{m^{\prime}l'}}
 \left.
 -\frac{3}{E_{ml}E_{m^{\prime }l'}E_{m^{\prime \prime }l}} -\frac{3}{E_{ml}E_{m^{\prime }l'}E_{m^{\prime \prime}l'}} \right).\end{align}%
This again can be simplified by approximating all the energy denominators as $-1$, to yield the last term of Eq.~(\ref{Heff_Vseries}). It should be clear that, neglecting the $O(1/\eta)$ corrections of the denominators, the generic form of the $n$-th order is:
\begin{equation}
\langle m | H_{eff}^{(2n)} | m^{\prime }\rangle  \simeq \alpha_n {\sum} \langle m | V_{od} | l^{(n)} \rangle   \langle l^{(n)} |V_{od} | m^{(n)} \rangle
\ldots
 \langle m^{\prime\prime} | V_{od} | l' \rangle \langle l' |V_{od} | m^{\prime} \rangle =\alpha_n\left\langle m|V_{\rm od}^{2n}|m^{\prime
}\right\rangle.
\end{equation}
where the coefficient $\alpha_n$ depends on the detailed form of the perturbation theory formula at order $n$. As a consequence, we have:
\begin{equation}\label{Heff_Vseries2}
H_{eff} \simeq \frac{p^{2}+x^2}{2\eta}- \sum_{n=0}^\infty \alpha_n V^{2n} .
\end{equation}

To find the values of $\alpha_n$, the simplest way is to consider the auxiliary problem with $H_0=\frac12 \sigma_z$ and $V_{od}= \epsilon \sigma_x$. The advantage of this Hamiltonian is that, in applying the perturbation theory formulas, the denominators are exactly equal to $-1$. The exact energy of the ground state is $-\sqrt{1/4+\epsilon^2}$ which can be expanded in Taylor series and allows to extract $\alpha_n = (-2)^{n-1}(2n-3)!!/n! $. This expression is in agreement with the first few values $\alpha_{0}=1/2$, $\alpha_1=1$ and $\alpha_2 = -1$, known from Eq.~(\ref{Heff_Vseries}).

Actually, the ground state energy of $\frac12 \sigma_z +\ \epsilon \sigma_x$ also indicates that the closed form of the series in Eq.~(\ref{Heff_Vseries2}) is simply $\sqrt{1/4+V^2}$. Thus, we obtain the final result of this section:
\begin{equation}\label{Heff_resummed}
H_{eff} \simeq \frac{p^{2}+x^2}{2\eta}- \sqrt{\frac{1}{4} +\frac{ \tilde{g}^2}{8\eta} \left[(1+\lambda)^2 x^2 +(1-\lambda)^2 p^2 -(1-\lambda^2)\right] }.
\end{equation}

\section{Expectation values in the superradiant phase}

In this section we compute several relevant expectation values in the $\eta \to \infty$ limit. In particular, we consider $\langle x^2 \rangle$, $\langle p^2 \rangle$,  $\langle x \sigma_z \rangle$, and $\langle p \sigma_y \rangle$, where the average is over the ground state of the anisotropic QRM. The behavior of these quantities at large $\eta$ justifies our discussion in the main text in terms of a ``Classical oscillator limit''.

Focusing on the x-type superradiant phase, we can make use of the effective Hamiltonian Eq.~(\ref{Heff_resummed}) to find the ground state. Following the discussion after Eq.~(10) of the main text, we have that $H_{eff}$ can be written as follows in the superradiant phase:
\begin{equation}\label{Heff_superradiant}
H_{eff} \simeq \frac{2 \lambda}{(1+\lambda)^2} \frac{p^2}{\eta}+ \frac{1}{2}\left(\frac{x^2}{\eta} - \sqrt{1 + 2\xi^2 \frac{x^2}{\eta}}\right),
\end{equation}
where the effective mass corresponds to Eq.~(10) of the main text, and the potential corresponds to $\tilde E_-$ [see Eq.~(6) of the main text]. In the superradiant phase, the potential has two minima which become equivalent when $\eta \to \infty$. For simplicity we focus on the one at $x_0 \simeq \sqrt{(\xi^2+\xi^{-2})\eta/2}$. Performing an harmonic approximation (valid for $\eta \to \infty$) we have:
\begin{equation}\label{Heff_harmonic}
H_{eff} \simeq \frac{2 \lambda}{\eta (1+\lambda)^2}p^2+\frac{\xi^4-1}{2\eta \xi^4}(x-x_0)^2,
\end{equation}
which allows us to immediately write the ground state $|\psi_0 \rangle$ for $H_{eff}$. In fact, remembering that $H_{eff}$ is valid in the $\sigma_x = -1$ subspace, the ground state is actually $|\Phi_0 \rangle = |\psi_0 \rangle |-\rangle$.

Given an operateor $A$, the expectation value $\langle \Phi_0  | A | \Phi_0 \rangle$ is easily computed. However, one should pay attention that $H_{eff}$ is obtained from the original $H$ after a unitary rotation $e^S$, thus the ground-state expectation value for the Rabi model is actually given by:
\begin{equation}
\langle A \rangle =  \langle \Phi_0  | e^S A e^{-S} | \Phi_0 \rangle.
\end{equation}
We then must consider first the derivation of suitable expressions for $S$ and $e^S$.

As it is standard in the Schrieffer-Wolff transformation, we expand $S$ in a series $S=\sum_n S_n$, where explicit expressions for the low-order $S_n$ are readily available (see, e.g., Ref.~\cite{Winkler_book}). Due to the structure of $S_n$, one can use an argument similar to the one we have developed for $H_{eff}$. Within the same approximation of Eq.~(\ref{Heff_resummed}), it easy to see that:
\begin{equation}
\langle l | S_{2n+1} | m \rangle  \simeq \beta_n {\sum} \langle l | V_{od} | m^{(n+1)} \rangle \langle m^{(n+1)} | V_{od} | l^{(n)} \rangle
\ldots
 \langle m^{\prime\prime} | V_{od} | l' \rangle \langle l' |V_{od} | m^{\prime} \rangle
 =\beta_n  \langle l |  V_{od}^{2n+1} | m \rangle ,
\end{equation}
while $S_{2n}=0$. To determine the coefficients $\beta_n$, we can consider again the auxiliary Hamiltonian $\frac12 \sigma_z +\ \epsilon \sigma_x$, for which $\langle \uparrow |S| \downarrow \rangle = 1/2 \arctan(2\epsilon)$. This leads to:
$
\langle l | S| m \rangle \simeq  1/2 \langle l |  \arctan(2 V_{od}) | m  \rangle =  1/2 \langle l | \sigma_x \arctan(2 V_{od}) | m\rangle$. The last step, i.e., inserting the  $\sigma_x$ operator, is justified because $| l \rangle$ is a $+1$ eigenstate of $\sigma_x$. This step allows us to write the matrix elements in terms of an anti-hermitian operator, as appropriate for $S$. Based on this we conclude:
\begin{equation}
S \simeq   \frac12  \sigma_x \arctan(2 V_{od}) = \frac12  \sigma_x \ \arctan{\left[\sqrt{\frac{2}{\eta}}\left(\xi x\sigma _{z}+\xi' p\sigma _{y}\right)\right]} ,
\end{equation}
which is consistent with $S_1$ given in Eq.~(\ref{S1}).

To proceed further, we write $S$ as a leading-order term plus a small $O(\eta^{-1/2})$ correction:
\begin{equation}
S \simeq    -i\frac{\sigma_y}{2}\arctan\sqrt{\xi^4-1}-i\frac{\xi^{-3}}{\sqrt{2\eta}} (x-x_0)\sigma_y +  i \frac{\arctan\sqrt{\xi^4-1}}{\sqrt{\xi^4-1}}\frac{\xi'}{\sqrt{2\eta}}p\sigma _{z}+ \ldots
\end{equation}
This expression allows us to approximate the unitary transformation to the same order. We obtain:
\begin{equation} \label{eS}
e^S   \simeq
e^{-i\frac{\sigma_y}{2}\arctan\sqrt{\xi^4-1}}
\left( 1-i\xi^{-3}\frac{x-x_0}{\sqrt{2\eta}} \sigma_y  \right)
+i\frac{\xi'/\xi}{\sqrt{\xi^2+1}}\frac{p}{\sqrt{\eta}}  \sigma_z + \ldots,
\end{equation}
which can be directly used to compute the leading-order of the various expectation values.

In fact, in several cases the approximation $e^S   \simeq   e^{-i\frac{\sigma_y}{2}\arctan\sqrt{\xi^4-1}}$ is sufficient. This includes  the evaluation of $\langle x^2 \rangle $ and $\langle p^2 \rangle$. Since the leading order of $e^S$ commutes with $x^2,~p^2$, we have:
\begin{align}
& \langle x^2 \rangle \simeq  \langle \Phi_0  |  x^2  | \Phi_0 \rangle \simeq x_0^2 = \frac{\xi^2+\xi^{-2}}{2}\eta, \\
& \langle p^2 \rangle \simeq  \langle \Phi_0  |  p^2  | \Phi_0 \rangle \simeq \frac{1+\lambda}{4\sqrt{\lambda}} \frac{\sqrt{\xi^4-1}}{\xi^2},
\end{align}
where we have made use of Eq.~(\ref{Heff_harmonic}). These results are in agreement with those cited in the main text. Using $e^S   \simeq   e^{-i\frac{\sigma_y}{2}\arctan\sqrt{\xi^4-1}}$ we can also compute $\langle x \sigma_z \rangle $:
\begin{equation}
\langle x \sigma_z \rangle  \simeq  \langle \Phi_0  | e^S  x \sigma_z e^{-S} | \Phi_0 \rangle \simeq x_0  \langle -  |  e^{-i\sigma_y \arctan\sqrt{\xi^4-1}}  \sigma_z | - \rangle =  -\sqrt{\frac{\eta}{2}} \left( \xi - \xi^{-3}\right).
\end{equation}
Finally, we consider $\langle p \sigma_y \rangle$, for which taking $e^S   \simeq   e^{-i\frac{\sigma_y}{2}\arctan\sqrt{\xi^4-1}}$ gives a vanishing result. In this case, we have to use the full expression given in Eq.~(\ref{eS}). A straightforward calculation gives:
\begin{equation}
\langle p \sigma_y \rangle = \frac{\xi^{-3}}{\sqrt{2\eta}}  \left( 1- \frac{1-\lambda}{2\sqrt{\lambda}}\sqrt{\xi^4-1}\right),
\end{equation}
which is the result cited in the main text. We have also confirmed that these analytical expressions reproduce  accurately the expectation values obtained by direct numerics at large $\eta$.

\section{Analytical consideration on the finite-$\eta$ scaling behavior}

We consider here the scaling form of the expectation values $\langle x^{2n}\rangle$ and $\langle p^{2n}\rangle$, at $\lambda \neq 0$ as well as $\lambda=0$ (the scaling behavior of the JC model was not discussed in Ref.~\cite{Hwang2016}). For $\lambda > 0$ we can immediately generalize the argument of the previous section to write:
\begin{equation}
 \langle x^{2n} \rangle \simeq  \langle \phi_0  |  x^{2n}  | \phi_0 \rangle, \qquad
 \langle p^{2n} \rangle \simeq  \langle \phi_0  |  p^{2n}  | \phi_0 \rangle ,
\end{equation}
where $|\phi_0 \rangle$ is the ground state of Eq.~(\ref{Heff_superradiant}) at $\xi \simeq 1$. As discussed in the main text, the wavefinction has the scaling form $\phi_0(x M _{\lambda}^{1/6}/\sqrt{\eta} ,tM_{\lambda}^{1/3})$ and after a simple change of variable the expectation values are written as:
\begin{align}
&\langle x^{2n} \rangle =\left( \eta M _{\lambda}^{-1/3} \right)^{n}
\left[\frac{ \int_{-\infty}^{\infty} \left|\phi_0 (u ,tM_{\lambda}^{1/3})\right|^2 u^{2n} du}
{ \int_{-\infty}^{\infty} \left|\phi_0 (u,tM_{\lambda}^{1/3})\right|^2 du}\right] , \\
&\langle p^{2n} \rangle =\left( \eta M _{\lambda}^{-1/3} \right)^{-n}
\left[\frac{ \int_{-\infty}^{\infty} \left | \frac{\partial^{n}}{\partial u^{n}}\phi_0(u ,tM_{\lambda}^{1/3}) \right|^2 du}
{ \int_{-\infty}^{\infty} \left|\phi_0(u,tM_{\lambda}^{1/3})\right|^2 du}\right],
\end{align}
where the expressions in the square parentheses are the explicit form for $X_n(tM_{\lambda}^{1/3})$, $P_n(tM_{\lambda}^{1/3})$ of the main text. For $\lambda<0$, using the exact mapping, we readily get the following scaling behavior (with $t'=\xi'-1$):
\begin{equation}
\langle x^{2n}\rangle=\left( \eta M _{-\lambda}^{-1/3} \right)^{-n} P_n(t'M_{-\lambda}^{1/3}), \qquad \langle p^{2n}\rangle=\left( \eta M _{-\lambda}^{-1/3} \right)^{n} X_n(t'M_{-\lambda}^{1/3}). \qquad\quad \rm{( for~} \lambda<0 {\rm )}
\end{equation}

Next, we consider the singular behavior at $\lambda =0$, where the well-known solution of the JC model allows for a relatively straighforward treatment. There are two eigenstates of the form $\alpha_q |q \rangle |-\rangle+\beta_q |q-1 \rangle |+\rangle$ ($q=1,2,\ldots$) and the one with lower energy gives:
\begin{equation}\label{Heff_JC}
E_{JC}(q)\simeq \frac{q}{\eta} - \frac{1}{2} \sqrt{\left(1-\frac{1}{\eta}\right)^2+ 4\xi^2 \frac{q}{\eta} }.\end{equation}
At $\xi<1$ the ground state is $|0\rangle|-\rangle$ while at $\xi>1$  the ground state is given by the $q_0$ which minimizes Eq.~(\ref{Heff_JC}).

As a side remark, we note that that Eq~(\ref{Heff_JC}) has the same form of $\tilde{E}_-$ if we approximate $1-1/\eta \simeq 1$ [see Eq.~(6) of the main text]. The only difference is that if $\tilde x^2$ is substituted here by $2q/\eta$. Therefore, by taking the $\eta \to \infty$ limit, $q_0/\eta$ is given by one half of Eq.~(7) of the main text. It is also easy to check that $\beta_q \to 0$ when $\eta \to \infty$, thus the ground state of the JC model approaches $|q_0 \rangle |-\rangle $. One immediate consequence, illustrated in Fig.~3(a) of the main text, is that  $\langle \tilde{x}^2 \rangle \simeq \langle q_0 | \tilde{x}^2 | q_0 \rangle \simeq q_0/\eta$ is exactly half of the value which one would obtain for $\lambda > 0$.

Returning to the derivation of the scaling function, we should determine the values $\xi_q$ at which the ground state changes from $|q\rangle$ to $|q+1 \rangle$. These are easily found from $E_{JC}(q)=E_{JC}(q+1)$ which, to leading order in $\eta$, gives:
\begin{equation}
\xi_q  \simeq 1+ \frac{q}{\eta}.
\end{equation}
Thus the $\xi_q$ are equally spaced on the $\xi$ axis, as illustrated in Fig.~3(b) of the main text. In each interval $[\xi_{q-1},\xi_q]$, the expectation values $\langle q |x^{2n} |q\rangle = \langle q |p^{2n} |q\rangle$ lead to Eq.~(15) of the main text.

\begin{figure}
\includegraphics[width=0.32\columnwidth]{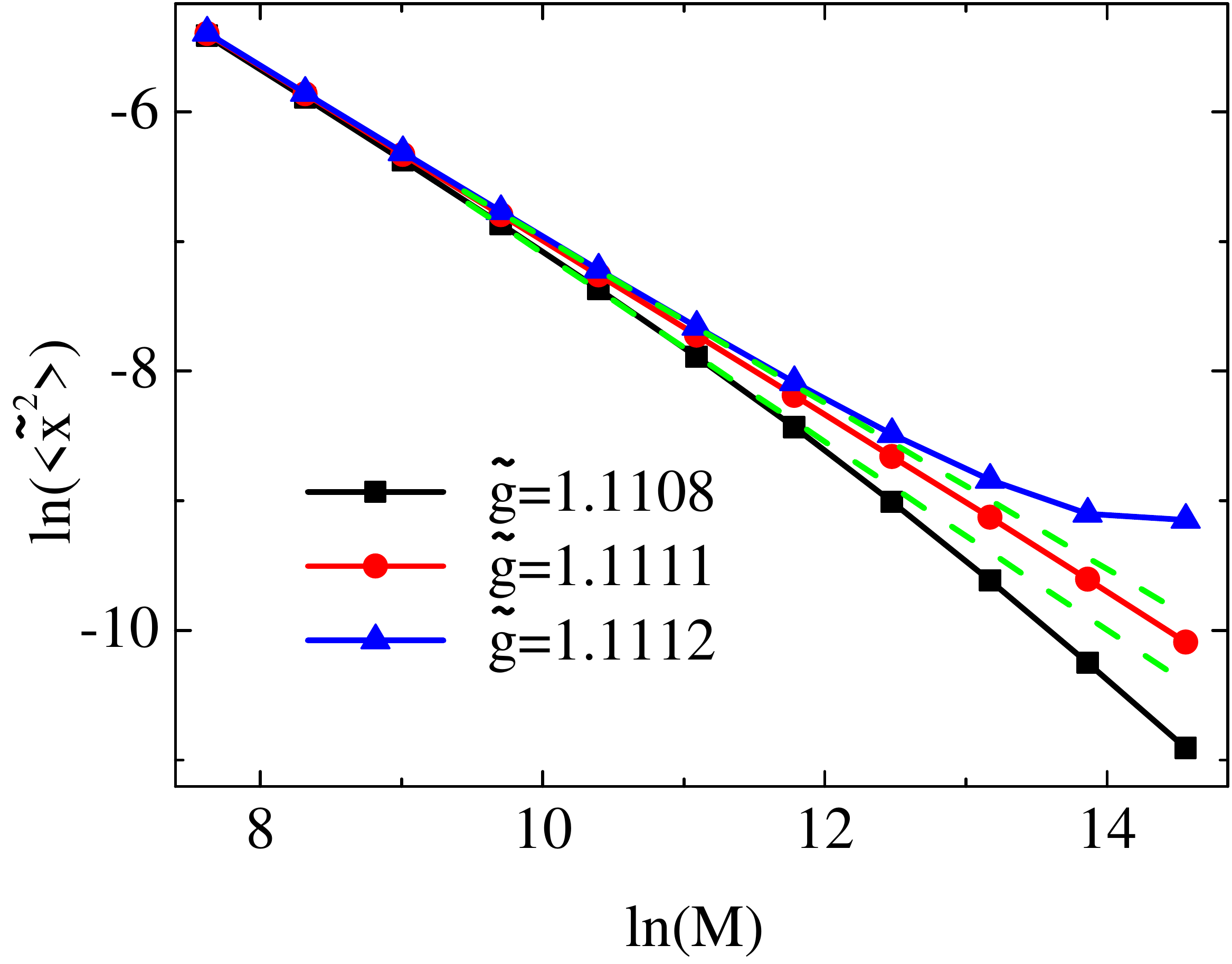}
\includegraphics[width=0.32\columnwidth]{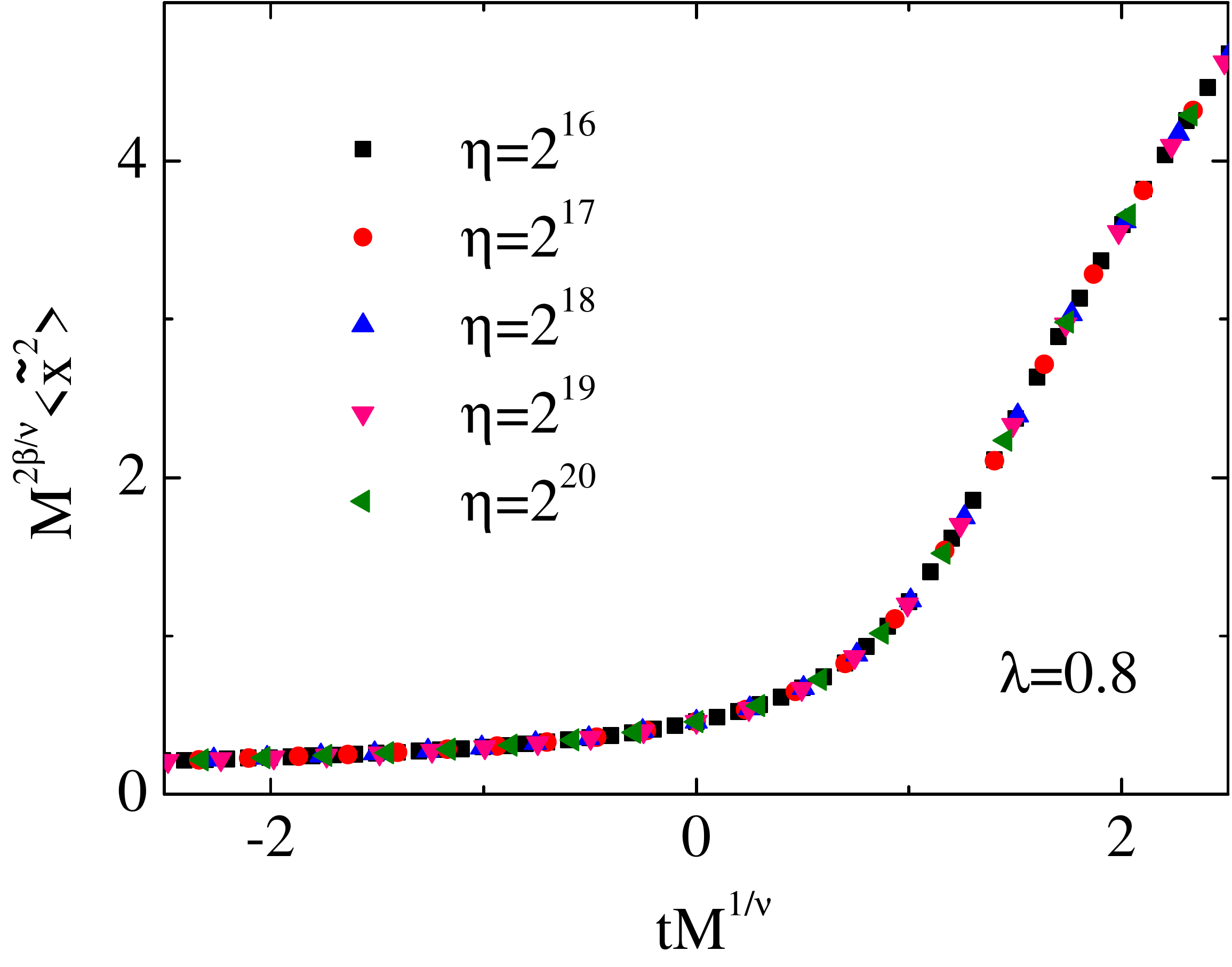}
\includegraphics[width=0.32\columnwidth]{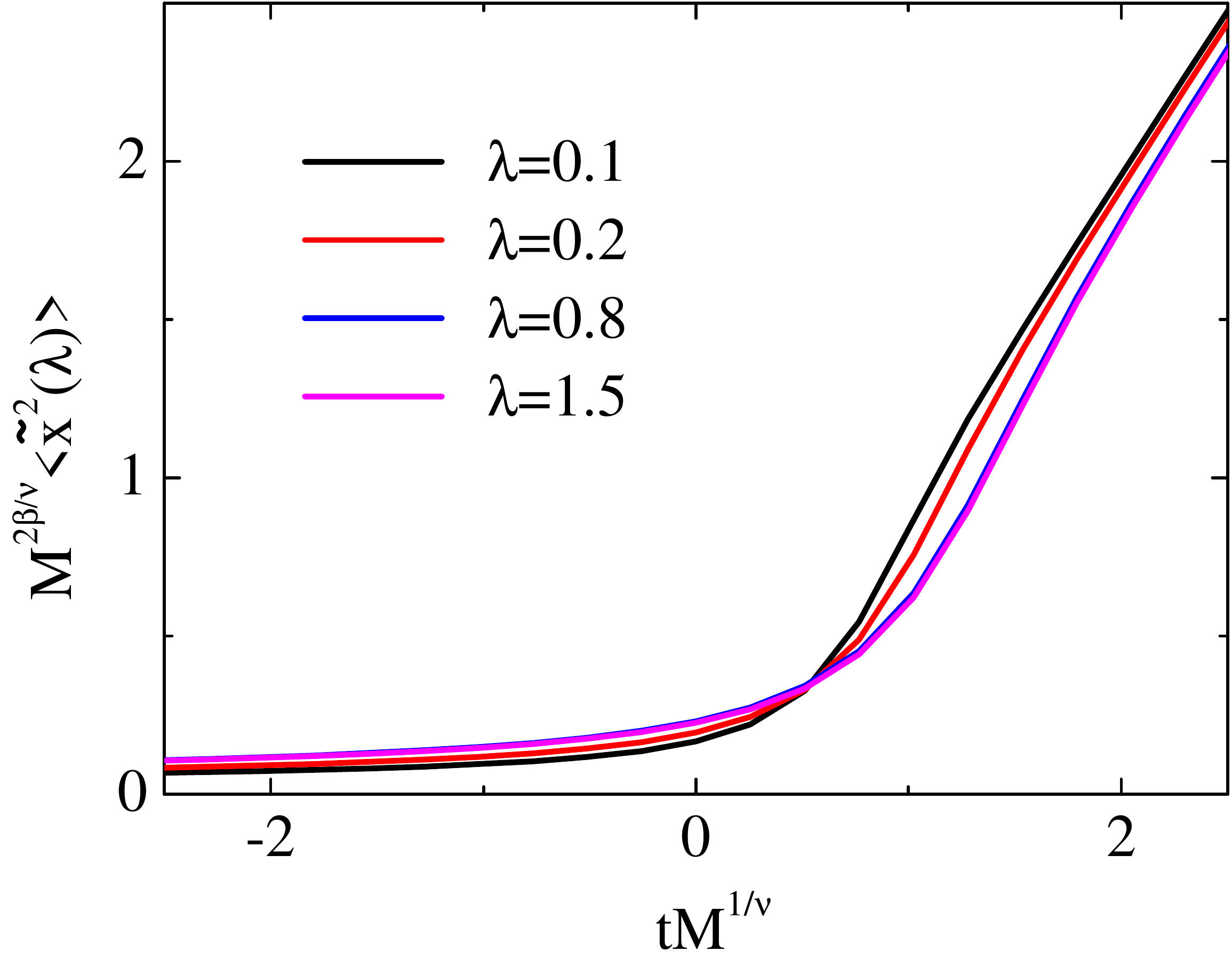}
\caption{(Color online) (a) Log-log plot of the $ \langle \tilde{x}^2\rangle $  for $\lambda=0.8$. The line at $\tilde{g}=1.1111$ is straight, indicating the critical point. Its slope is $-0.68 \pm 0.02$, giving the value of the critical exponent ratio $-2\beta/\nu$. The green straight dashed lines are guides to the eye, to clearly show that the lines below and above the critical coupling have a finite curvature. (b) The finite scaling function of  $\tilde{x}^2$ at $\lambda=0.8$. To obtain this scaling function we have chosen $1/\nu=0.33$.  (c) The finite-$\eta$ scaling function of $\tilde{x}^2$ at different values of $\lambda$, i.e., $\lambda=0.1$, 0.2, 0.8, and 1.5. As seen, using $M$ instead of $M_\lambda$ leads to distinct scaling functions.}
\label{num_slope}
\end{figure}

\section{Numerical scaling analysis}

In the following we describe a numerical analysis  of the scaling behavior. Having noticed the second-order nature of the phase transition at $\tilde{g}_c$, it is natural to speculate the following scaling law for a physical quantity $Q$ in the critical region:
\begin{equation}\label{scaling_Q_guessed}
Q(M, t ; \lambda)=M^{-\beta_Q/\nu}\tilde{Q}_{\lambda}(tM^{1/\nu}),
\end{equation}
where $M=\eta^2$ is large enough but finite, $t=(\tilde{g}-\tilde{g_c})/\tilde{g_c}$ is the reduced coupling, $\beta_{Q}$ is the critical exponent for $Q$, $\nu$ is the critical exponent for $M$, and $\tilde{Q}_{\lambda}$ is the scaling function of $Q$ at a given $\lambda$. This scaling form is in the same spirit of the finite size scaling in classical thermodynamics phase transitions. Notice also that, differently from the main text, we take here a blind numerical approach, i.e., we neglect the $\lambda$ renormalization of $M$ revealed by our analytical study. It is also convenient to take the logarithm on both side of Eq.~\eqref{scaling_Q_guessed}, to obtain the following $\log$-$\log$ relation between $\ln Q$ and $\ln M$:
\begin{equation}\label{log_log_Q}
\ln Q(M, t ; \lambda)=\frac{\beta_Q}{\nu}\ln M+\ln \tilde{Q}_{\lambda}(tM^{1/\nu}).
\end{equation}
At critical point $t=0$, there is linear relation between $\ln Q$ and $\ln M$:
\begin{equation}\label{linearQ}
\ln Q=\frac{\beta_Q}{\nu}\ln M+\ln \tilde{Q}(0).
\end{equation}
where the slope of the linear dependence is the critical exponent ratio $\beta_Q/\nu$. Based on Eq.~\eqref{linearQ}, we can determine the critical point as well as the critical exponent ratio $\beta_Q/\nu$.

We illustrate this procedure by considering $Q=\langle \tilde{x}^2 \rangle$ at $\lambda=0.8$ as an example, for which we define:
\begin{equation}
\langle \tilde{x}^2 \rangle= M^{-2\beta/\nu}X^2_{\lambda}(tM^{1/\nu}).
\end{equation}
The left panel of Fig.~\ref{num_slope} shows the $\log$-$\log$ relation with coupling strengths around the critical value $\tilde{g}_c=1.1111$. At the critical coupling $\tilde{g_c}$, we can see that the curve of $\ln \langle \tilde{x}^2 \rangle$ as a function of $\ln M$ becomes a straight line, confirming the linear relation of Eq.~\eqref{linearQ}.  For small deviations from the critical coupling (i.e., $\tilde{g}=1.1108$ and $\tilde{g}=1.1112$), the linear character of $\ln \langle \tilde{x}^2\rangle$ breaks down. As seen in the first panel of Fig.~\ref{num_slope}, the curves at both sides of the critical point have a finite and opposite curvature. Thus, we can estimate the critical coupling as $\tilde{g}_c =1.1111\pm0.0003$. At the same time, the slope of the straight line at $\tilde{g}=\tilde{g}_c$ gives the critical exponent ratio $\beta/\nu=0.33$. Furthermore, to confirm the scaling relation of the type Eq.~\eqref{scaling_Q_guessed} and determine the critical exponent $\nu$, we can set $v=tM^{1/\nu}$ as the variable of the  horizontal axis and $M^{2\beta/\nu}\langle \tilde{x}^2 \rangle$ as the variable of the vertical axis. The middle panel of Fig.~\ref{num_slope} shows that, if the value of $\nu$ is properly chosen, curves with different scales of $M$ collapse into a single curve which corresponds to the scaling function of $\tilde{Q}_{\lambda}$. Doing this, we obtain $1/\nu=0.33\pm0.02$.

\begin{table*}
\begin{tabular}{|c|c|c|c|c|c|c|c|c|c|c|c|c|c|}
\hline
\hline
$\lambda$  & $0.1$ & $0.2$ & $0.3$ & $0.4$ & $0.5$ & $0.6$ & $0.7$ & $0.8$ & $0.9$ & $1.0$ & $1.5$ & $2.0$ & $5.0$  \\
\hline
\hline
$\tilde{g}_c$ (Analytical)& $20/11$ & $5/3$ & $20/13$ & $10/7$ & $3/2$ &   $4/3$ & $20/17$&      $10/9$ & $20/19$& $1$ & $4/5$ & $2/3$ & $1/3$ \\
\hline
$\tilde{g}_c$ (Numerical)& $1.18182$ & $1.6667$ & $1.5384$ & $1.4286$ & $1.4999$ & $1.3334$& $1.1765$ & $1.1111$ & $1.0526$ & $1.0002$ &  $0.7999$ & $0.6666$ & $0.3334$\\
\hline
$\beta/\nu$ (Analytical)& $1/3$ & $1/3$ & $1/3$ & $1/3$ & $1/3$ & $1/3$& $1/3$ & $1/3$ & $1/3$ & $1/3$ & $1/3$ & $1/3$&     $1/3$\\
\hline
$\beta/\nu$ (Numerical)& $0.34$ & $0.33$ & $0.34$ & $0.32$ & $0.33$ & $0.33$& $0.35$ & $0.33$ & $0.32$ & $0.33$ & $0.34$ & $0.33$&      $0.33$\\
\hline
$1/\nu$ (Analytical)& $1/3$ & $1/3$ & $1/3$ & $1/3$ & $1/3$ & $1/3$& $1/3$ & $1/3$ & $1/3$ & $1/3$ & $1/3$ & $1/3$&    $1/3$ \\
\hline
$1/\nu$ (Numerical)& $0.33$ & $0.32$ & $0.33$ & $0.32$ & $0.34$ & $0.33$& $0.34$ & $0.33$ & $0.33$ & $0.33$ & $0.32$ & $0.33$&      $0.33$\\
\hline
\hline
\end{tabular}
\caption{Critical point $\tilde{g_c}$, critical exponent ratio $\beta/\nu$, and critical exponent $1/\nu$ at different values of the anisotropy parameter $\lambda$. Both numerical and analytical results are presented. In the numerical calculations, the error for $\tilde{g}_c$ is less than $0.0005$ and the error for $\beta/\nu$ ($1/\nu$) is less than $0.02$. }
\label{table}
\end{table*}

We have performed this numerical analysis for several values of $\lambda$ and collected the results in Table~\ref{table}, which also shows a comparison to the critical couplings and exponents obtained by our analytical method.  We see that the two methods are in good agreement. The advantage of the numeirical approach is that it can be always applied, even when an analytical treatment might be difficult to achieve (i.e., for other more complicated models). However, the analytical approach allows here to reach several important conclusions regarding the universlity of the phase transition. First, the analytical results shows that the critical exponents are exactly identical for different values of $\lambda$ (while the numerical method can only establish the equality within errors). Second, the renormalization of $M$ into $M_\lambda$ is not obvious from the direct numerical scaling. To make this point explicit we show in the right panel of Fig.~\ref{num_slope} that, within this traditional framework, the scaling functions at different $\lambda$ are actually different. Only by substituting $M$ with the renormalized value $M_{\lambda}$  the scaling functions at different $\lambda$ become identical, as shown in Fig.~2(a) of the main text. To establish that two critical points belong to  the same universality class, it is widely acknowledged that one should prove the identity of both critical exponents and scaling functions. Thus, the analytical results are  instrumental here to conclude that models with different values of $\lambda$ belong to the same universality class.

\end{widetext}


\begin{thebibliography}{37}%
\makeatletter
\providecommand \@ifxundefined [1]{%
 \@ifx{#1\undefined}
}%
\providecommand \@ifnum [1]{%
 \ifnum #1\expandafter \@firstoftwo
 \else \expandafter \@secondoftwo
 \fi
}%
\providecommand \@ifx [1]{%
 \ifx #1\expandafter \@firstoftwo
 \else \expandafter \@secondoftwo
 \fi
}%
\providecommand \natexlab [1]{#1}%
\providecommand \enquote  [1]{``#1''}%
\providecommand \bibnamefont  [1]{#1}%
\providecommand \bibfnamefont [1]{#1}%
\providecommand \citenamefont [1]{#1}%
\providecommand \href@noop [0]{\@secondoftwo}%
\providecommand \href [0]{\begingroup \@sanitize@url \@href}%
\providecommand \@href[1]{\@@startlink{#1}\@@href}%
\providecommand \@@href[1]{\endgroup#1\@@endlink}%
\providecommand \@sanitize@url [0]{\catcode `\\12\catcode `\$12\catcode
  `\&12\catcode `\#12\catcode `\^12\catcode `\_12\catcode `\%12\relax}%
\providecommand \@@startlink[1]{}%
\providecommand \@@endlink[0]{}%
\providecommand \url  [0]{\begingroup\@sanitize@url \@url }%
\providecommand \@url [1]{\endgroup\@href {#1}{\urlprefix }}%
\providecommand \urlprefix  [0]{URL }%
\providecommand \Eprint [0]{\href }%
\providecommand \doibase [0]{http://dx.doi.org/}%
\providecommand \selectlanguage [0]{\@gobble}%
\providecommand \bibinfo  [0]{\@secondoftwo}%
\providecommand \bibfield  [0]{\@secondoftwo}%
\providecommand \translation [1]{[#1]}%
\providecommand \BibitemOpen [0]{}%
\providecommand \bibitemStop [0]{}%
\providecommand \bibitemNoStop [0]{.\EOS\space}%
\providecommand \EOS [0]{\spacefactor3000\relax}%
\providecommand \BibitemShut  [1]{\csname bibitem#1\endcsname}%
\let\auto@bib@innerbib\@empty
\bibitem [{\citenamefont {Hwang}\ \emph {et~al.}(2015)\citenamefont {Hwang},
  \citenamefont {Puebla},\ and\ \citenamefont {Plenio}}]{Hwang2015}%
  \BibitemOpen
  \bibfield  {author} {\bibinfo {author} {\bibfnamefont {M.-J.}\ \bibnamefont
  {Hwang}}, \bibinfo {author} {\bibfnamefont {R.}~\bibnamefont {Puebla}}, \
  and\ \bibinfo {author} {\bibfnamefont {M.~B.}\ \bibnamefont {Plenio}},\
  }\href {\doibase 10.1103/PhysRevLett.115.180404} {\bibfield  {journal}
  {\bibinfo  {journal} {Phys. Rev. Lett.}\ }\textbf {\bibinfo {volume} {115}},\
  \bibinfo {pages} {180404} (\bibinfo {year} {2015})}\BibitemShut {NoStop}%
\bibitem [{\citenamefont {Hwang}\ and\ \citenamefont
  {Plenio}(2016)}]{Hwang2016}%
  \BibitemOpen
  \bibfield  {author} {\bibinfo {author} {\bibfnamefont {M.-J.}\ \bibnamefont
  {Hwang}}\ and\ \bibinfo {author} {\bibfnamefont {M.~B.}\ \bibnamefont
  {Plenio}},\ }\href {\doibase 10.1103/PhysRevLett.117.123602} {\bibfield
  {journal} {\bibinfo  {journal} {Phys. Rev. Lett.}\ }\textbf {\bibinfo
  {volume} {117}},\ \bibinfo {pages} {123602} (\bibinfo {year}
  {2016})}\BibitemShut {NoStop}%
\bibitem [{\citenamefont {Wallraff}\ \emph {et~al.}(2004)\citenamefont
  {Wallraff}, \citenamefont {Schuster}, \citenamefont {Blais}, \citenamefont
  {Frunzio}, \citenamefont {Huang}, \citenamefont {Majer}, \citenamefont
  {Kumar}, \citenamefont {Girvin},\ and\ \citenamefont
  {Schoelkopf}}]{Wallraff2004}%
  \BibitemOpen
  \bibfield  {author} {\bibinfo {author} {\bibfnamefont {A.~A.}\ \bibnamefont
  {Wallraff}}, \bibinfo {author} {\bibfnamefont {D.~I.}\ \bibnamefont
  {Schuster}}, \bibinfo {author} {\bibfnamefont {A.}~\bibnamefont {Blais}},
  \bibinfo {author} {\bibfnamefont {L.}~\bibnamefont {Frunzio}}, \bibinfo
  {author} {\bibfnamefont {R.-S.}\ \bibnamefont {Huang}}, \bibinfo {author}
  {\bibfnamefont {J.}~\bibnamefont {Majer}}, \bibinfo {author} {\bibfnamefont
  {S.}~\bibnamefont {Kumar}}, \bibinfo {author} {\bibfnamefont {S.~M.}\
  \bibnamefont {Girvin}}, \ and\ \bibinfo {author} {\bibfnamefont {R.~J.}\
  \bibnamefont {Schoelkopf}},\ }\href {\doibase 10.1038/nature02851} {\bibfield
   {journal} {\bibinfo  {journal} {Nature}\ }\textbf {\bibinfo {volume}
  {431}},\ \bibinfo {pages} {162} (\bibinfo {year} {2004})}\BibitemShut
  {NoStop}%
\bibitem [{\citenamefont {Devoret}\ \emph {et~al.}(2007)\citenamefont
  {Devoret}, \citenamefont {Girvin},\ and\ \citenamefont
  {Schoelkopf}}]{Devoret2007}%
  \BibitemOpen
  \bibfield  {author} {\bibinfo {author} {\bibfnamefont {M.}~\bibnamefont
  {Devoret}}, \bibinfo {author} {\bibfnamefont {S.}~\bibnamefont {Girvin}}, \
  and\ \bibinfo {author} {\bibfnamefont {R.}~\bibnamefont {Schoelkopf}},\
  }\href {\doibase 10.1002/andp.200710261} {\bibfield  {journal} {\bibinfo
  {journal} {Annalen der Physik}\ }\textbf {\bibinfo {volume} {16}},\ \bibinfo
  {pages} {767} (\bibinfo {year} {2007})}\BibitemShut {NoStop}%
\bibitem [{\citenamefont {Bourassa}\ \emph {et~al.}(2009)\citenamefont
  {Bourassa}, \citenamefont {Gambetta}, \citenamefont {Abdumalikov},
  \citenamefont {Astafiev}, \citenamefont {Nakamura},\ and\ \citenamefont
  {Blais}}]{Bourassa2009}%
  \BibitemOpen
  \bibfield  {author} {\bibinfo {author} {\bibfnamefont {J.}~\bibnamefont
  {Bourassa}}, \bibinfo {author} {\bibfnamefont {J.~M.}\ \bibnamefont
  {Gambetta}}, \bibinfo {author} {\bibfnamefont {A.~A.}\ \bibnamefont
  {Abdumalikov}}, \bibinfo {author} {\bibfnamefont {O.}~\bibnamefont
  {Astafiev}}, \bibinfo {author} {\bibfnamefont {Y.}~\bibnamefont {Nakamura}},
  \ and\ \bibinfo {author} {\bibfnamefont {A.}~\bibnamefont {Blais}},\ }\href
  {\doibase 10.1103/PhysRevA.80.032109} {\bibfield  {journal} {\bibinfo
  {journal} {Phys. Rev. A}\ }\textbf {\bibinfo {volume} {80}},\ \bibinfo
  {pages} {032109} (\bibinfo {year} {2009})}\BibitemShut {NoStop}%
\bibitem [{\citenamefont {Niemczyk}\ \emph {et~al.}(2010)\citenamefont
  {Niemczyk}, \citenamefont {Deppe}, \citenamefont {Huebl}, \citenamefont
  {Menzel}, \citenamefont {Hocke}, \citenamefont {Schwarz}, \citenamefont
  {Garcia-Ripoll}, \citenamefont {Zueco}, \citenamefont {Hummer}, \citenamefont
  {Solano}, \citenamefont {Marx},\ and\ \citenamefont {Gross}}]{Niemczyk2010}%
  \BibitemOpen
  \bibfield  {author} {\bibinfo {author} {\bibfnamefont {T.~A.}\ \bibnamefont
  {Niemczyk}}, \bibinfo {author} {\bibfnamefont {F.}~\bibnamefont {Deppe}},
  \bibinfo {author} {\bibfnamefont {H.}~\bibnamefont {Huebl}}, \bibinfo
  {author} {\bibfnamefont {E.~P.}\ \bibnamefont {Menzel}}, \bibinfo {author}
  {\bibfnamefont {F.}~\bibnamefont {Hocke}}, \bibinfo {author} {\bibfnamefont
  {M.~J.}\ \bibnamefont {Schwarz}}, \bibinfo {author} {\bibfnamefont {J.~J.}\
  \bibnamefont {Garcia-Ripoll}}, \bibinfo {author} {\bibfnamefont
  {D.}~\bibnamefont {Zueco}}, \bibinfo {author} {\bibfnamefont
  {T.}~\bibnamefont {Hummer}}, \bibinfo {author} {\bibfnamefont
  {E.}~\bibnamefont {Solano}}, \bibinfo {author} {\bibfnamefont
  {A.}~\bibnamefont {Marx}}, \ and\ \bibinfo {author} {\bibfnamefont
  {R.}~\bibnamefont {Gross}},\ }\href {\doibase 10.1038/nphys1730} {\bibfield
  {journal} {\bibinfo  {journal} {Nat. Phys.}\ }\textbf {\bibinfo {volume}
  {6}},\ \bibinfo {pages} {772} (\bibinfo {year} {2010})}\BibitemShut {NoStop}%
\bibitem [{\citenamefont {Peropadre}\ \emph {et~al.}(2010)\citenamefont
  {Peropadre}, \citenamefont {Forn-D\'{\i}az}, \citenamefont {Solano},\ and\
  \citenamefont {Garc\'{\i}a-Ripoll}}]{Peropadre2010}%
  \BibitemOpen
  \bibfield  {author} {\bibinfo {author} {\bibfnamefont {B.}~\bibnamefont
  {Peropadre}}, \bibinfo {author} {\bibfnamefont {P.}~\bibnamefont
  {Forn-D\'{\i}az}}, \bibinfo {author} {\bibfnamefont {E.}~\bibnamefont
  {Solano}}, \ and\ \bibinfo {author} {\bibfnamefont {J.~J.}\ \bibnamefont
  {Garc\'{\i}a-Ripoll}},\ }\href {\doibase 10.1103/PhysRevLett.105.023601}
  {\bibfield  {journal} {\bibinfo  {journal} {Phys. Rev. Lett.}\ }\textbf
  {\bibinfo {volume} {105}},\ \bibinfo {pages} {023601} (\bibinfo {year}
  {2010})}\BibitemShut {NoStop}%
\bibitem [{\citenamefont {Ridolfo}\ \emph {et~al.}(2012)\citenamefont
  {Ridolfo}, \citenamefont {Leib}, \citenamefont {Savasta},\ and\ \citenamefont
  {Hartmann}}]{Ridolfo2012}%
  \BibitemOpen
  \bibfield  {author} {\bibinfo {author} {\bibfnamefont {A.}~\bibnamefont
  {Ridolfo}}, \bibinfo {author} {\bibfnamefont {M.}~\bibnamefont {Leib}},
  \bibinfo {author} {\bibfnamefont {S.}~\bibnamefont {Savasta}}, \ and\
  \bibinfo {author} {\bibfnamefont {M.~J.}\ \bibnamefont {Hartmann}},\ }\href
  {\doibase 10.1103/PhysRevLett.109.193602} {\bibfield  {journal} {\bibinfo
  {journal} {Phys. Rev. Lett.}\ }\textbf {\bibinfo {volume} {109}},\ \bibinfo
  {pages} {193602} (\bibinfo {year} {2012})}\BibitemShut {NoStop}%
\bibitem [{\citenamefont {Todorov}\ and\ \citenamefont
  {Sirtori}(2014)}]{Todorov2014}%
  \BibitemOpen
  \bibfield  {author} {\bibinfo {author} {\bibfnamefont {Y.}~\bibnamefont
  {Todorov}}\ and\ \bibinfo {author} {\bibfnamefont {C.}~\bibnamefont
  {Sirtori}},\ }\href {\doibase 10.1103/PhysRevX.4.041031} {\bibfield
  {journal} {\bibinfo  {journal} {Phys. Rev. X}\ }\textbf {\bibinfo {volume}
  {4}},\ \bibinfo {pages} {041031} (\bibinfo {year} {2014})}\BibitemShut
  {NoStop}%
\bibitem [{\citenamefont {Baust}\ \emph {et~al.}(2016)\citenamefont {Baust},
  \citenamefont {Hoffmann}, \citenamefont {Haeberlein}, \citenamefont
  {Schwarz}, \citenamefont {Eder}, \citenamefont {Goetz}, \citenamefont
  {Wulschner}, \citenamefont {Xie}, \citenamefont {Zhong}, \citenamefont
  {Quijandr\'{\i}a}, \citenamefont {Zueco}, \citenamefont {Ripoll},
  \citenamefont {Garc\'{\i}a-\'Alvarez}, \citenamefont {Romero}, \citenamefont
  {Solano}, \citenamefont {Fedorov}, \citenamefont {Menzel}, \citenamefont
  {Deppe}, \citenamefont {Marx},\ and\ \citenamefont {Gross}}]{Baust2016}%
  \BibitemOpen
  \bibfield  {author} {\bibinfo {author} {\bibfnamefont {A.}~\bibnamefont
  {Baust}}, \bibinfo {author} {\bibfnamefont {E.}~\bibnamefont {Hoffmann}},
  \bibinfo {author} {\bibfnamefont {M.}~\bibnamefont {Haeberlein}}, \bibinfo
  {author} {\bibfnamefont {M.~J.}\ \bibnamefont {Schwarz}}, \bibinfo {author}
  {\bibfnamefont {P.}~\bibnamefont {Eder}}, \bibinfo {author} {\bibfnamefont
  {J.}~\bibnamefont {Goetz}}, \bibinfo {author} {\bibfnamefont
  {F.}~\bibnamefont {Wulschner}}, \bibinfo {author} {\bibfnamefont
  {E.}~\bibnamefont {Xie}}, \bibinfo {author} {\bibfnamefont {L.}~\bibnamefont
  {Zhong}}, \bibinfo {author} {\bibfnamefont {F.}~\bibnamefont
  {Quijandr\'{\i}a}}, \bibinfo {author} {\bibfnamefont {D.}~\bibnamefont
  {Zueco}}, \bibinfo {author} {\bibfnamefont {J.-J.~G.}\ \bibnamefont
  {Ripoll}}, \bibinfo {author} {\bibfnamefont {L.}~\bibnamefont
  {Garc\'{\i}a-\'Alvarez}}, \bibinfo {author} {\bibfnamefont {G.}~\bibnamefont
  {Romero}}, \bibinfo {author} {\bibfnamefont {E.}~\bibnamefont {Solano}},
  \bibinfo {author} {\bibfnamefont {K.~G.}\ \bibnamefont {Fedorov}}, \bibinfo
  {author} {\bibfnamefont {E.~P.}\ \bibnamefont {Menzel}}, \bibinfo {author}
  {\bibfnamefont {F.}~\bibnamefont {Deppe}}, \bibinfo {author} {\bibfnamefont
  {A.}~\bibnamefont {Marx}}, \ and\ \bibinfo {author} {\bibfnamefont
  {R.}~\bibnamefont {Gross}},\ }\href {\doibase 10.1103/PhysRevB.93.214501}
  {\bibfield  {journal} {\bibinfo  {journal} {Phys. Rev. B}\ }\textbf {\bibinfo
  {volume} {93}},\ \bibinfo {pages} {214501} (\bibinfo {year}
  {2016})}\BibitemShut {NoStop}%
\bibitem [{\citenamefont {Forn-Diaz}\ \emph {et~al.}(2016)\citenamefont
  {Forn-Diaz}, \citenamefont {Romero}, \citenamefont {Harmans}, \citenamefont
  {Solano},\ and\ \citenamefont {Mooij}}]{Forn-Diaz2016}%
  \BibitemOpen
  \bibfield  {author} {\bibinfo {author} {\bibfnamefont {P.}~\bibnamefont
  {Forn-Diaz}}, \bibinfo {author} {\bibfnamefont {G.}~\bibnamefont {Romero}},
  \bibinfo {author} {\bibfnamefont {C.~J. P.~M.}\ \bibnamefont {Harmans}},
  \bibinfo {author} {\bibfnamefont {E.}~\bibnamefont {Solano}}, \ and\ \bibinfo
  {author} {\bibfnamefont {J.~E.}\ \bibnamefont {Mooij}},\ }\href {\doibase
  10.1038/srep26720} {\bibfield  {journal} {\bibinfo  {journal} {Sci. Rep.}\
  }\textbf {\bibinfo {volume} {6}},\ \bibinfo {pages} {26720} (\bibinfo {year}
  {2016})}\BibitemShut {NoStop}%
\bibitem [{\citenamefont {Casanova}\ \emph {et~al.}(2010)\citenamefont
  {Casanova}, \citenamefont {Romero}, \citenamefont {Lizuain}, \citenamefont
  {Garc\'{\i}a-Ripoll},\ and\ \citenamefont {Solano}}]{Casanova2010}%
  \BibitemOpen
  \bibfield  {author} {\bibinfo {author} {\bibfnamefont {J.}~\bibnamefont
  {Casanova}}, \bibinfo {author} {\bibfnamefont {G.}~\bibnamefont {Romero}},
  \bibinfo {author} {\bibfnamefont {I.}~\bibnamefont {Lizuain}}, \bibinfo
  {author} {\bibfnamefont {J.~J.}\ \bibnamefont {Garc\'{\i}a-Ripoll}}, \ and\
  \bibinfo {author} {\bibfnamefont {E.}~\bibnamefont {Solano}},\ }\href
  {\doibase 10.1103/PhysRevLett.105.263603} {\bibfield  {journal} {\bibinfo
  {journal} {Phys. Rev. Lett.}\ }\textbf {\bibinfo {volume} {105}},\ \bibinfo
  {pages} {263603} (\bibinfo {year} {2010})}\BibitemShut {NoStop}%
\bibitem [{\citenamefont {De~Liberato}(2014)}]{Liberato2014}%
  \BibitemOpen
  \bibfield  {author} {\bibinfo {author} {\bibfnamefont {S.}~\bibnamefont
  {De~Liberato}},\ }\href {\doibase 10.1103/PhysRevLett.112.016401} {\bibfield
  {journal} {\bibinfo  {journal} {Phys. Rev. Lett.}\ }\textbf {\bibinfo
  {volume} {112}},\ \bibinfo {pages} {016401} (\bibinfo {year}
  {2014})}\BibitemShut {NoStop}%
\bibitem [{\citenamefont {Fornd¨ªaz}\ \emph {et~al.}(2017)\citenamefont
  {Fornd¨ªaz}, \citenamefont {Garc¨ªaripoll}, \citenamefont {Peropadre},
  \citenamefont {Orgiazzi}, \citenamefont {Yurtalan}, \citenamefont
  {Belyansky}, \citenamefont {Wilson},\ and\ \citenamefont
  {Lupascu}}]{Fornd2017}%
  \BibitemOpen
  \bibfield  {author} {\bibinfo {author} {\bibfnamefont {P.}~\bibnamefont
  {Fornd¨ªaz}}, \bibinfo {author} {\bibfnamefont {J.~J.}\ \bibnamefont
  {Garc¨ªaripoll}}, \bibinfo {author} {\bibfnamefont {B.}~\bibnamefont
  {Peropadre}}, \bibinfo {author} {\bibfnamefont {J.~L.}\ \bibnamefont
  {Orgiazzi}}, \bibinfo {author} {\bibfnamefont {M.~A.}\ \bibnamefont
  {Yurtalan}}, \bibinfo {author} {\bibfnamefont {R.}~\bibnamefont {Belyansky}},
  \bibinfo {author} {\bibfnamefont {C.~M.}\ \bibnamefont {Wilson}}, \ and\
  \bibinfo {author} {\bibfnamefont {A.}~\bibnamefont {Lupascu}},\ }\href@noop
  {} {\bibfield  {journal} {\bibinfo  {journal} {Nat. Phys.}\ }\textbf
  {\bibinfo {volume} {13}},\ \bibinfo {pages} {39} (\bibinfo {year}
  {2017})}\BibitemShut {NoStop}%
\bibitem [{\citenamefont {{Yoshihara}}\ \emph {et~al.}(2017)\citenamefont
  {{Yoshihara}}, \citenamefont {{Fuse}}, \citenamefont {{Ashhab}},
  \citenamefont {{Kakuyanagi}}, \citenamefont {{Saito}},\ and\ \citenamefont
  {{Semba}}}]{Yoshihara2017}%
  \BibitemOpen
  \bibfield  {author} {\bibinfo {author} {\bibfnamefont {F.}~\bibnamefont
  {{Yoshihara}}}, \bibinfo {author} {\bibfnamefont {T.}~\bibnamefont {{Fuse}}},
  \bibinfo {author} {\bibfnamefont {S.}~\bibnamefont {{Ashhab}}}, \bibinfo
  {author} {\bibfnamefont {K.}~\bibnamefont {{Kakuyanagi}}}, \bibinfo {author}
  {\bibfnamefont {S.}~\bibnamefont {{Saito}}}, \ and\ \bibinfo {author}
  {\bibfnamefont {K.}~\bibnamefont {{Semba}}},\ }\href@noop {} {\bibfield
  {journal} {\bibinfo  {journal} {Nat. Phys.}\ }\textbf {\bibinfo {volume}
  {13}},\ \bibinfo {pages} {44} (\bibinfo {year} {2017})}\BibitemShut {NoStop}%
\bibitem [{\citenamefont {Chen}\ \emph {et~al.}(2016)\citenamefont {Chen},
  \citenamefont {Wang}, \citenamefont {Li}, \citenamefont {Tian}, \citenamefont
  {Qiu}, \citenamefont {Inomata}, \citenamefont {Yoshihara}, \citenamefont
  {Han}, \citenamefont {Nori}, \citenamefont {Tsai},\ and\ \citenamefont
  {You}}]{chenzhen2017}%
  \BibitemOpen
  \bibfield  {author} {\bibinfo {author} {\bibfnamefont {Z.}~\bibnamefont
  {Chen}}, \bibinfo {author} {\bibfnamefont {Y.}~\bibnamefont {Wang}}, \bibinfo
  {author} {\bibfnamefont {T.}~\bibnamefont {Li}}, \bibinfo {author}
  {\bibfnamefont {L.}~\bibnamefont {Tian}}, \bibinfo {author} {\bibfnamefont
  {Y.}~\bibnamefont {Qiu}}, \bibinfo {author} {\bibfnamefont {K.}~\bibnamefont
  {Inomata}}, \bibinfo {author} {\bibfnamefont {F.}~\bibnamefont {Yoshihara}},
  \bibinfo {author} {\bibfnamefont {S.}~\bibnamefont {Han}}, \bibinfo {author}
  {\bibfnamefont {F.}~\bibnamefont {Nori}}, \bibinfo {author} {\bibfnamefont
  {J.~S.}\ \bibnamefont {Tsai}}, \ and\ \bibinfo {author} {\bibfnamefont
  {J.~Q.}\ \bibnamefont {You}},\ }\href@noop {} {\bibfield  {journal} {\bibinfo
   {journal} {arXiv:1602.01584v1}\ } (\bibinfo {year} {2016})}\BibitemShut
  {NoStop}%
\bibitem [{\citenamefont {Rabi}(1936)}]{Rabi1936}%
  \BibitemOpen
  \bibfield  {author} {\bibinfo {author} {\bibfnamefont {I.~I.}\ \bibnamefont
  {Rabi}},\ }\href {\doibase 10.1103/PhysRev.49.324} {\bibfield  {journal}
  {\bibinfo  {journal} {Phys. Rev.}\ }\textbf {\bibinfo {volume} {49}},\
  \bibinfo {pages} {324} (\bibinfo {year} {1936})}\BibitemShut {NoStop}%
\bibitem [{\citenamefont {Rabi}(1937)}]{Rabi1937}%
  \BibitemOpen
  \bibfield  {author} {\bibinfo {author} {\bibfnamefont {I.~I.}\ \bibnamefont
  {Rabi}},\ }\href {\doibase 10.1103/PhysRev.51.652} {\bibfield  {journal}
  {\bibinfo  {journal} {Phys. Rev.}\ }\textbf {\bibinfo {volume} {51}},\
  \bibinfo {pages} {652} (\bibinfo {year} {1937})}\BibitemShut {NoStop}%
\bibitem [{\citenamefont {Forn-D\'{\i}az}\ \emph {et~al.}(2010)\citenamefont
  {Forn-D\'{\i}az}, \citenamefont {Lisenfeld}, \citenamefont {Marcos},
  \citenamefont {Garc\'{\i}a-Ripoll}, \citenamefont {Solano}, \citenamefont
  {Harmans},\ and\ \citenamefont {Mooij}}]{Forn-Diaz2010}%
  \BibitemOpen
  \bibfield  {author} {\bibinfo {author} {\bibfnamefont {P.}~\bibnamefont
  {Forn-D\'{\i}az}}, \bibinfo {author} {\bibfnamefont {J.}~\bibnamefont
  {Lisenfeld}}, \bibinfo {author} {\bibfnamefont {D.}~\bibnamefont {Marcos}},
  \bibinfo {author} {\bibfnamefont {J.~J.}\ \bibnamefont {Garc\'{\i}a-Ripoll}},
  \bibinfo {author} {\bibfnamefont {E.}~\bibnamefont {Solano}}, \bibinfo
  {author} {\bibfnamefont {C.~J. P.~M.}\ \bibnamefont {Harmans}}, \ and\
  \bibinfo {author} {\bibfnamefont {J.~E.}\ \bibnamefont {Mooij}},\ }\href
  {\doibase 10.1103/PhysRevLett.105.237001} {\bibfield  {journal} {\bibinfo
  {journal} {Phys. Rev. Lett.}\ }\textbf {\bibinfo {volume} {105}},\ \bibinfo
  {pages} {237001} (\bibinfo {year} {2010})}\BibitemShut {NoStop}%
\bibitem [{\citenamefont {Braak}(2011)}]{Braak2011}%
  \BibitemOpen
  \bibfield  {author} {\bibinfo {author} {\bibfnamefont {D.}~\bibnamefont
  {Braak}},\ }\href {\doibase 10.1103/PhysRevLett.107.100401} {\bibfield
  {journal} {\bibinfo  {journal} {Phys. Rev. Lett.}\ }\textbf {\bibinfo
  {volume} {107}},\ \bibinfo {pages} {100401} (\bibinfo {year}
  {2011})}\BibitemShut {NoStop}%
\bibitem [{\citenamefont {Solano}(2011)}]{Solano2011}%
  \BibitemOpen
  \bibfield  {author} {\bibinfo {author} {\bibfnamefont {E.}~\bibnamefont
  {Solano}},\ }\href {http://physics.aps.org/articles/v4/68} {\bibfield
  {journal} {\bibinfo  {journal} {Physics}\ }\textbf {\bibinfo {volume} {4}},\
  \bibinfo {pages} {68} (\bibinfo {year} {2011})}\BibitemShut {NoStop}%
\bibitem [{\citenamefont {Bakemeier}\ \emph {et~al.}(2012)\citenamefont
  {Bakemeier}, \citenamefont {Alvermann},\ and\ \citenamefont
  {Fehske}}]{Bakemeier2012}%
  \BibitemOpen
  \bibfield  {author} {\bibinfo {author} {\bibfnamefont {L.}~\bibnamefont
  {Bakemeier}}, \bibinfo {author} {\bibfnamefont {A.}~\bibnamefont
  {Alvermann}}, \ and\ \bibinfo {author} {\bibfnamefont {H.}~\bibnamefont
  {Fehske}},\ }\href {\doibase 10.1103/PhysRevA.85.043821} {\bibfield
  {journal} {\bibinfo  {journal} {Phys. Rev. A}\ }\textbf {\bibinfo {volume}
  {85}},\ \bibinfo {pages} {043821} (\bibinfo {year} {2012})}\BibitemShut
  {NoStop}%
\bibitem [{\citenamefont {Chen}\ \emph {et~al.}(2012)\citenamefont {Chen},
  \citenamefont {Wang}, \citenamefont {He}, \citenamefont {Liu},\ and\
  \citenamefont {Wang}}]{Chen2012}%
  \BibitemOpen
  \bibfield  {author} {\bibinfo {author} {\bibfnamefont {Q.-H.}\ \bibnamefont
  {Chen}}, \bibinfo {author} {\bibfnamefont {C.}~\bibnamefont {Wang}}, \bibinfo
  {author} {\bibfnamefont {S.}~\bibnamefont {He}}, \bibinfo {author}
  {\bibfnamefont {T.}~\bibnamefont {Liu}}, \ and\ \bibinfo {author}
  {\bibfnamefont {K.-L.}\ \bibnamefont {Wang}},\ }\href {\doibase
  10.1103/PhysRevA.86.023822} {\bibfield  {journal} {\bibinfo  {journal} {Phys.
  Rev. A}\ }\textbf {\bibinfo {volume} {86}},\ \bibinfo {pages} {023822}
  (\bibinfo {year} {2012})}\BibitemShut {NoStop}%
\bibitem [{\citenamefont {Ashhab}(2013)}]{Ashhab2013}%
  \BibitemOpen
  \bibfield  {author} {\bibinfo {author} {\bibfnamefont {S.}~\bibnamefont
  {Ashhab}},\ }\href {\doibase 10.1103/PhysRevA.87.013826} {\bibfield
  {journal} {\bibinfo  {journal} {Phys. Rev. A}\ }\textbf {\bibinfo {volume}
  {87}},\ \bibinfo {pages} {013826} (\bibinfo {year} {2013})}\BibitemShut
  {NoStop}%
\bibitem [{\citenamefont {Wolf}\ \emph {et~al.}(2013)\citenamefont {Wolf},
  \citenamefont {Vallone}, \citenamefont {Romero}, \citenamefont {Kollar},
  \citenamefont {Solano},\ and\ \citenamefont {Braak}}]{Wolf2013}%
  \BibitemOpen
  \bibfield  {author} {\bibinfo {author} {\bibfnamefont {F.~A.}\ \bibnamefont
  {Wolf}}, \bibinfo {author} {\bibfnamefont {F.}~\bibnamefont {Vallone}},
  \bibinfo {author} {\bibfnamefont {G.}~\bibnamefont {Romero}}, \bibinfo
  {author} {\bibfnamefont {M.}~\bibnamefont {Kollar}}, \bibinfo {author}
  {\bibfnamefont {E.}~\bibnamefont {Solano}}, \ and\ \bibinfo {author}
  {\bibfnamefont {D.}~\bibnamefont {Braak}},\ }\href {\doibase
  10.1103/PhysRevA.87.023835} {\bibfield  {journal} {\bibinfo  {journal} {Phys.
  Rev. A}\ }\textbf {\bibinfo {volume} {87}},\ \bibinfo {pages} {023835}
  (\bibinfo {year} {2013})}\BibitemShut {NoStop}%
\bibitem [{\citenamefont {Ying}\ \emph {et~al.}(2015)\citenamefont {Ying},
  \citenamefont {Liu}, \citenamefont {Luo}, \citenamefont {Lin},\ and\
  \citenamefont {You}}]{Ying2015}%
  \BibitemOpen
  \bibfield  {author} {\bibinfo {author} {\bibfnamefont {Z.-J.}\ \bibnamefont
  {Ying}}, \bibinfo {author} {\bibfnamefont {M.}~\bibnamefont {Liu}}, \bibinfo
  {author} {\bibfnamefont {H.-G.}\ \bibnamefont {Luo}}, \bibinfo {author}
  {\bibfnamefont {H.-Q.}\ \bibnamefont {Lin}}, \ and\ \bibinfo {author}
  {\bibfnamefont {J.~Q.}\ \bibnamefont {You}},\ }\href {\doibase
  10.1103/PhysRevA.92.053823} {\bibfield  {journal} {\bibinfo  {journal} {Phys.
  Rev. A}\ }\textbf {\bibinfo {volume} {92}},\ \bibinfo {pages} {053823}
  (\bibinfo {year} {2015})}\BibitemShut {NoStop}%
\bibitem [{\citenamefont {Braak}\ \emph {et~al.}(2016)\citenamefont {Braak},
  \citenamefont {Chen}, \citenamefont {Batchelor},\ and\ \citenamefont
  {Solano}}]{Braak2016}%
  \BibitemOpen
  \bibfield  {author} {\bibinfo {author} {\bibfnamefont {D.}~\bibnamefont
  {Braak}}, \bibinfo {author} {\bibfnamefont {Q.-H.}\ \bibnamefont {Chen}},
  \bibinfo {author} {\bibfnamefont {M.~T.}\ \bibnamefont {Batchelor}}, \ and\
  \bibinfo {author} {\bibfnamefont {E.}~\bibnamefont {Solano}},\ }\href
  {http://stacks.iop.org/1751-8121/49/i=30/a=300301} {\bibfield  {journal}
  {\bibinfo  {journal} {Journal of Physics A: Mathematical and Theoretical}\
  }\textbf {\bibinfo {volume} {49}},\ \bibinfo {pages} {300301} (\bibinfo
  {year} {2016})}\BibitemShut {NoStop}%
\bibitem [{\citenamefont {Jaynes}\ and\ \citenamefont
  {Cummings}(1963)}]{Jaynes1963}%
  \BibitemOpen
  \bibfield  {author} {\bibinfo {author} {\bibfnamefont {E.~T.}\ \bibnamefont
  {Jaynes}}\ and\ \bibinfo {author} {\bibfnamefont {F.~W.}\ \bibnamefont
  {Cummings}},\ }\href {\doibase 10.1109/PROC.1963.1664} {\bibfield  {journal}
  {\bibinfo  {journal} {Proceedings of the IEEE}\ }\textbf {\bibinfo {volume}
  {51}},\ \bibinfo {pages} {89} (\bibinfo {year} {1963})}\BibitemShut {NoStop}%
\bibitem [{\citenamefont {Xie}\ \emph {et~al.}(2014)\citenamefont {Xie},
  \citenamefont {Cui}, \citenamefont {Cao}, \citenamefont {Amico},\ and\
  \citenamefont {Fan}}]{Xie2014}%
  \BibitemOpen
  \bibfield  {author} {\bibinfo {author} {\bibfnamefont {Q.-T.}\ \bibnamefont
  {Xie}}, \bibinfo {author} {\bibfnamefont {S.}~\bibnamefont {Cui}}, \bibinfo
  {author} {\bibfnamefont {J.-P.}\ \bibnamefont {Cao}}, \bibinfo {author}
  {\bibfnamefont {L.}~\bibnamefont {Amico}}, \ and\ \bibinfo {author}
  {\bibfnamefont {H.}~\bibnamefont {Fan}},\ }\href {\doibase
  10.1103/PhysRevX.4.021046} {\bibfield  {journal} {\bibinfo  {journal} {Phys.
  Rev. X}\ }\textbf {\bibinfo {volume} {4}},\ \bibinfo {pages} {021046}
  (\bibinfo {year} {2014})}\BibitemShut {NoStop}%
\bibitem [{\citenamefont {Schliemann}\ \emph {et~al.}(2002)\citenamefont
  {Schliemann}, \citenamefont {Egues},\ and\ \citenamefont
  {Loss}}]{Schliemann2003}%
  \BibitemOpen
  \bibfield  {author} {\bibinfo {author} {\bibfnamefont {J.}~\bibnamefont
  {Schliemann}}, \bibinfo {author} {\bibfnamefont {J.~C.}\ \bibnamefont
  {Egues}}, \ and\ \bibinfo {author} {\bibfnamefont {D.}~\bibnamefont {Loss}},\
  }\href@noop {} {\bibfield  {journal} {\bibinfo  {journal} {Phys. Rev. B}\
  }\textbf {\bibinfo {volume} {67}},\ \bibinfo {pages} {085302} (\bibinfo
  {year} {2002})}\BibitemShut {NoStop}%
\bibitem [{\citenamefont {Wang}\ \emph {et~al.}(2016)\citenamefont {Wang},
  \citenamefont {Zheng}, \citenamefont {Wang},\ and\ \citenamefont
  {Li}}]{Wangzh2016}%
  \BibitemOpen
  \bibfield  {author} {\bibinfo {author} {\bibfnamefont {Z.~H.}\ \bibnamefont
  {Wang}}, \bibinfo {author} {\bibfnamefont {Q.}~\bibnamefont {Zheng}},
  \bibinfo {author} {\bibfnamefont {X.}~\bibnamefont {Wang}}, \ and\ \bibinfo
  {author} {\bibfnamefont {Y.}~\bibnamefont {Li}},\ }\href@noop {} {\bibfield
  {journal} {\bibinfo  {journal} {Sci. Rep.}\ }\textbf {\bibinfo {volume}
  {6}},\ \bibinfo {pages} {22347} (\bibinfo {year} {2016})}\BibitemShut
  {NoStop}%
\bibitem [{\citenamefont {Baksic}\ and\ \citenamefont
  {Ciuti}(2014)}]{Baksic2014}%
  \BibitemOpen
  \bibfield  {author} {\bibinfo {author} {\bibfnamefont {A.}~\bibnamefont
  {Baksic}}\ and\ \bibinfo {author} {\bibfnamefont {C.}~\bibnamefont {Ciuti}},\
  }\href {\doibase 10.1103/PhysRevLett.112.173601} {\bibfield  {journal}
  {\bibinfo  {journal} {Phys. Rev. Lett.}\ }\textbf {\bibinfo {volume} {112}},\
  \bibinfo {pages} {173601} (\bibinfo {year} {2014})}\BibitemShut {NoStop}%
\bibitem [{\citenamefont {Puebla}\ \emph {et~al.}(2016)\citenamefont {Puebla},
  \citenamefont {Casanova},\ and\ \citenamefont {Plenio}}]{Puebla2016}%
  \BibitemOpen
  \bibfield  {author} {\bibinfo {author} {\bibfnamefont {R.}~\bibnamefont
  {Puebla}}, \bibinfo {author} {\bibfnamefont {J.}~\bibnamefont {Casanova}}, \
  and\ \bibinfo {author} {\bibfnamefont {M.~B.}\ \bibnamefont {Plenio}},\
  }\href@noop {} {\bibfield  {journal} {\bibinfo  {journal} {New J. Phys.}\
  }\textbf {\bibinfo {volume} {18}},\ \bibinfo {pages} {113039} (\bibinfo
  {year} {2016})}\BibitemShut {NoStop}%
\bibitem [{Sup()}]{Suppl_Info}%
  \BibitemOpen
  \href@noop {} {}\bibinfo {note} {See the Supplemental Information for the
  derivation of useful effective Hamiltonians, expectation values, and the
  scaling functions of the JC model, as well as further details on the
  numerical scaling analysis.}\BibitemShut {Stop}%
\bibitem [{\citenamefont {Larson}\ and\ \citenamefont
  {Irish}(2016)}]{Jonas2016}%
  \BibitemOpen
  \bibfield  {author} {\bibinfo {author} {\bibfnamefont {J.}~\bibnamefont
  {Larson}}\ and\ \bibinfo {author} {\bibfnamefont {E.~K.}\ \bibnamefont
  {Irish}},\ }\href@noop {} {\bibfield  {journal} {\bibinfo  {journal}
  {arXiv:1612.00336}\ } (\bibinfo {year} {2016})}\BibitemShut {NoStop}%
\bibitem [{\citenamefont {Gr{\"o}blacher}\ \emph {et~al.}(2009)\citenamefont
  {Gr{\"o}blacher}, \citenamefont {Hammerer}, \citenamefont {Vanner},\ and\
  \citenamefont {Aspelmeyer}}]{Groblacher2009}%
  \BibitemOpen
  \bibfield  {author} {\bibinfo {author} {\bibfnamefont {S.}~\bibnamefont
  {Gr{\"o}blacher}}, \bibinfo {author} {\bibfnamefont {K.}~\bibnamefont
  {Hammerer}}, \bibinfo {author} {\bibfnamefont {M.~R.}\ \bibnamefont
  {Vanner}}, \ and\ \bibinfo {author} {\bibfnamefont {M.}~\bibnamefont
  {Aspelmeyer}},\ }\href@noop {} {\bibfield  {journal} {\bibinfo  {journal}
  {Nature}\ }\textbf {\bibinfo {volume} {460}},\ \bibinfo {pages} {724}
  (\bibinfo {year} {2009})}\BibitemShut {NoStop}%
\bibitem [{\citenamefont {G{\"u}nter}\ \emph {et~al.}(2009)\citenamefont
  {G{\"u}nter}, \citenamefont {Anappara}, \citenamefont {Hees}, \citenamefont
  {Sell}, \citenamefont {Biasiol}, \citenamefont {Sorba}, \citenamefont
  {De~Liberato}, \citenamefont {Ciuti}, \citenamefont {Tredicucci},
  \citenamefont {Leitenstorfer} \emph {et~al.}}]{Gunter2009}%
  \BibitemOpen
  \bibfield  {author} {\bibinfo {author} {\bibfnamefont {G.}~\bibnamefont
  {G{\"u}nter}}, \bibinfo {author} {\bibfnamefont {A.~A.}\ \bibnamefont
  {Anappara}}, \bibinfo {author} {\bibfnamefont {J.}~\bibnamefont {Hees}},
  \bibinfo {author} {\bibfnamefont {A.}~\bibnamefont {Sell}}, \bibinfo {author}
  {\bibfnamefont {G.}~\bibnamefont {Biasiol}}, \bibinfo {author} {\bibfnamefont
  {L.}~\bibnamefont {Sorba}}, \bibinfo {author} {\bibfnamefont
  {S.}~\bibnamefont {De~Liberato}}, \bibinfo {author} {\bibfnamefont
  {C.}~\bibnamefont {Ciuti}}, \bibinfo {author} {\bibfnamefont
  {A.}~\bibnamefont {Tredicucci}}, \bibinfo {author} {\bibfnamefont
  {A.}~\bibnamefont {Leitenstorfer}},  \emph {et~al.},\ }\href@noop {}
  {\bibfield  {journal} {\bibinfo  {journal} {Nature}\ }\textbf {\bibinfo
  {volume} {458}},\ \bibinfo {pages} {178} (\bibinfo {year}
  {2009})}\BibitemShut {NoStop}%
  
  \bibitem{SC}
F.~L.~Pedrocchi, S.~Chesi, and D.~Loss, Phys. Rev. B \textbf{83}, 115415 (2011).

\bibitem{Winkler_book} R. Winkler, \emph{Spin-Orbit Coupling Effects in Two-
Dimensional Electron and Hole Systems} (Springer, Berlin,
2003).
  
\end{thebibliography}
%

\end{document}